\documentclass[fleqn,usenatbib]{mnras}

\newcommand{\avg}[1]{\langle#1\rangle}
 
\newcommand{\beq}{\begin{equation}}
\newcommand{\eeq}{\end{equation}}
\newcommand{\bit}{\begin{itemize}}
\newcommand{\eit}{\end{itemize}}

\newcommand{\beqa}{\begin{equation}\begin{aligned}}
\newcommand{\eeqa}{\end{aligned}\end{equation}}
\newcommand{\comment}[1]{}

\newcommand{\lnM}{\ln M}

\newcommand{\DS}{\Delta\Sigma}

\newcommand{\Ngal}{N_{\rm gal}}

\usepackage{newtxtext,newtxmath}

\usepackage[T1]{fontenc}

\usepackage{lineno}

\DeclareRobustCommand{\VAN}[3]{#2}
\let\VANthebibliography\thebibliography
\def\thebibliography{\DeclareRobustCommand{\VAN}[3]{##3}\VANthebibliography}

\usepackage{epsfig}

\usepackage[dvipsnames]{xcolor}
\usepackage{color}
\usepackage{booktabs}
\newcommand{\ra}[1]{\renewcommand{\arraystretch}{#1}}
\usepackage{float}

\usepackage{amsmath}    

\usepackage{mathtools}  
\usepackage{amssymb}    

\usepackage{verbatim}   
\usepackage{hyperref}   
\usepackage{multicol}  %
\usepackage{multirow}
\usepackage{ifthen}  
\usepackage{graphicx}
\usepackage{caption}
\usepackage{subcaption}
\usepackage{lipsum}       
\usepackage{changepage}   
\usepackage{enumitem}
\usepackage{multirow}
\usepackage{threeparttable}

\newcommand{\kms}{{\rm \, km~s}\ensuremath{^{-1}}}
\newcommand{\DSig}{\Delta\Sigma}

\defcitealias{Evrard:2014}{E14}

\title[Impact of Property Covariance on WL Scaling Relations]{Impact of Property Covariance on Cluster Weak Lensing Scaling Relations}
\author[Zhang et al.]{Zhuowen Zhang$^{1}$\thanks{E-mail:zzhang13@uchicago.edu}, Arya Farahi$^{2}$, Daisuke Nagai$^{3}$, Erwin T.\ Lau$^{4}$, Joshua Frieman$^{1}$, Marina Ricci$^{5}$, \newauthor
 Anja von der Linden$^{6}$, Hao-Yi Wu$^{7}$, and the LSST Dark Energy Science Collaboration
\\
$^{1}$Kavli Institute for Cosmological Physics, University of Chicago, Chicago, IL 60637, USA\\
$^{2}$Departments of Statistics and Data Science, University of Texas at Austin, Austin, TX 78757, USA\\
$^{3}$Department of Physics, Yale University, New Haven, CT 06520, USA\\
$^{4}$Center for Astrophysics | Harvard \& Smithsonian, Cambridge, MA, 02138, USA\\
$^{5}$Université Paris Cité, CNRS, Astroparticule et Cosmologie, F-75013 Paris, France \\
$^{6}$Department of Physics and Astronomy, Stony Brooks University, Stony Brook, NY, 11794, USA \\
$^{7}$Department of Physics, Boise State University,  Boise, ID, 83725, USA 
}

\date{Accepted XXX. Received YYY; in original form ZZZ}

\pubyear{}

\begin{document}
\label{firstpage}
\pagerange{\pageref{firstpage}--\pageref{lastpage}}
\maketitle

\begin{abstract}
We present an investigation into a hitherto unexplored systematic that affects the accuracy of galaxy cluster mass estimates with weak gravitational lensing. Specifically, we study the covariance between the weak lensing signal, $\Delta\Sigma$, and the ``true'' cluster galaxy number count, $N_{\rm gal}$, as measured within a spherical volume that is void of projection effects. By quantifying the impact of this covariance on mass calibration, this work reveals a significant source of systematic uncertainty. Using the MDPL2 simulation with galaxies traced by the SAGE semi-analytic model, we measure the intrinsic property covariance between these observables within the 3D vicinity of the cluster, spanning a range of dynamical mass and redshift values relevant for optical cluster surveys. Our results reveal a negative covariance at small radial scales ($R \lesssim R_{\rm 200c}$) and a null covariance at large scales ($R \gtrsim R_{\rm 200c}$) across most mass and redshift bins. We also find that this covariance results in a $2-3\%$ bias in the halo mass estimates in most bins. Furthermore, by modeling $N_{\rm gal}$ and $\Delta\Sigma$ as multi-(log)-linear equations of secondary halo properties, we provide a quantitative explanation for the physical origin of the negative covariance at small scales. Specifically, we demonstrate that the $\Ngal$--$\DSig$ covariance can be explained by the secondary properties of halos that probe their formation history. We attribute the difference between our results and the positive bias seen in other works with (mock)-cluster finders to projection effects. These findings highlight the importance of accounting for the covariance between observables in cluster mass estimation, which is crucial for obtaining accurate constraints on cosmological parameters.
\end{abstract}




\section{Introduction}

\begin{figure*}
    \centering
    \includegraphics[width=\linewidth]{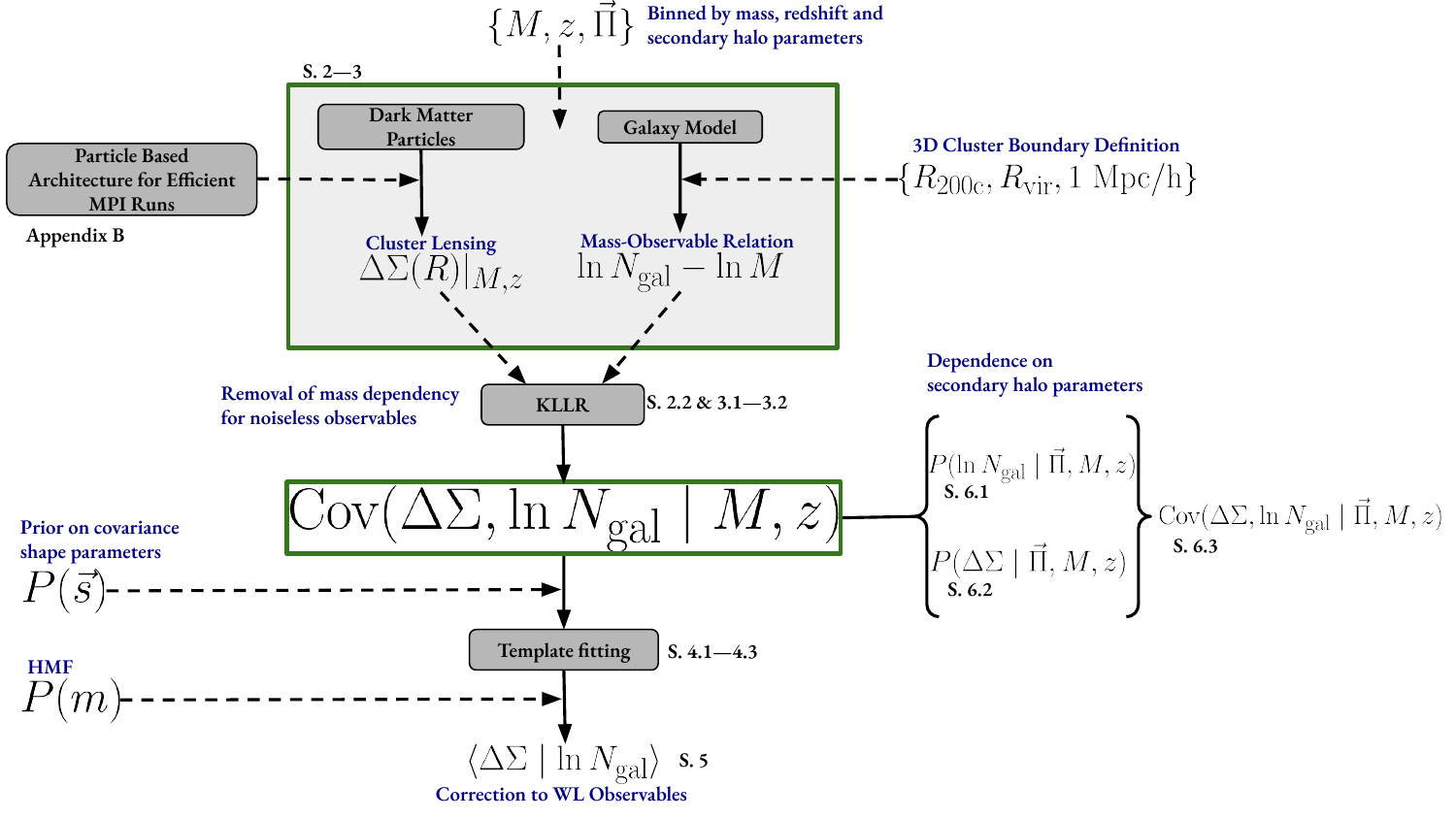}
    \caption{Graphic representation of the modeling of $\text{Cov}(\Delta\Sigma, \ln N_{\rm gal} \mid M, z)$ --- the covariance between the halo weak lensing signal $\Delta\Sigma(R)$ and log-richness $\ln N_{\rm gal}$ conditioned on mass and redshift --- and its dependence with secondary halo parameters $\Pi$. The labels marked \textit{S. XX} point to the location in the text. A full list of the notations used in this paper is introduced in Table \ref{tab:notation} \& \ref{tab:scaling_notation}.} 
    \label{fig:Outline}
\end{figure*}

Cluster abundance and its evolution with redshift are linked to the constituents of the Universe through the growth of cosmic structure. 
 \citep[][for a review]{Allen:2011}. 
Cluster abundance measured in large-scale galaxy surveys offers power constraints on cosmological parameters \citep[e.g.,][]{Vikhlinin2009,Mantz:2015,deHaan:2016,Mantz:2016,DES-DE+:2016,Pierre:2016,Costanzi_2021}.
These constraints are based on accurate cluster mass measurements, which are not directly observable and must be inferred. 
Cluster mass calibration has been identified as one of the leading systematic uncertainties in cosmological constraints using galaxy cluster abundance \citep[][]{Mantz2010, Rozo2010, Linden2014, Applegate2014, Dodelson2016,  Murata2019, Costanzi_2021}.
Accurate mappings between a population of massive clusters and their observables are thus critical and essential in cluster cosmology.  
Considerable effort has been put into measuring the statistical relationships between masses and observable properties that reflect their baryon contents \citep[see][ for a review]{Giodini:2013} and quantifying the sources of uncertainties.  

The Dark Energy Survey (DES) cluster cosmology from the Year 1 dataset \citep{Abbott_2020} reported tension in $\Omega_m$ --- the mean matter density of the universe --- with the DES 3x2pt probe that utilizes three two-point functions from the DES galaxy survey \citep{DES_3x2pt_Y1}. The tension between these two probes that utilize the same underlying dataset may be attributed to systematics that bias the weak lensing mass of clusters low at the low mass end \citep{To_2021, Costanzi_2021}. A possible origin for this discrepancy is that cluster masses are biased low due to systematics in cluster mass calibration. On the other hand, the tension can also originate from new physics that extends the Standard Cosmological model. Thus, it is important to understand the systematics of cluster mass calibration. Cluster masses estimated from X-ray and SZ data are known to suffer from hydrostatic bias \citep{Pratt:2019}. Conversely, cluster masses estimated from weak lensing have the potential to be more accurate compared to X-ray and SZ cluster masses. The systematics in the weak lensing mass calibration has just started to be explored recently \citep[][]{Applegate2014, Schrabback2018, McClintock2019,Kiiveri2021,Wu_2022}.

A relatively unexplored category of cluster systematics is the covariance between different cluster properties, including cluster observables and mass proxies. 
In cluster mass calibration, it is often assumed that this property covariance is negligible.  However, as initially pointed out by \citet{Nord:2008} and later shown in \citet[][]{Evrard:2014} and \citet[][]{Farahi2018}, non-zero property covariances between cluster observables can induce non-trivial, additive bias in cluster mass. As property covariance is additive, the systematic uncertainties that it induces will not be mitigated with the reduction of statistical errors as the sample size of the cluster increases.
To achieve accurate cosmological constraints with the next generation of large-scale cluster surveys, it is imperative that systematic uncertainties in the property covariance be accurately and precisely quantified \citep[][]{Rozo2014}. 

Although the property covariance linking mass to observable properties is becoming better understood and measured \citep{Wu:2015,Mantz:2016,Farahi2018,Farahi2019,Sereno2020}, studies that specifically investigate weak lensing property covariance are scarce, which poses a challenge for upcoming lensing surveys of galaxy clusters such as the Rubin \citep{LSST_2019} observatories. To achieve the percentage-level lensing mass calibration goals for the upcoming observations, the property covariance of weak lensing must be quantified.  

The physical origins of property covariance in lensing signals of galaxy clusters can be attributed to the halo formation history of the cluster and baryonic physics \citep[][]{Xhakaj2022}. Developing a first-principle physical model for the property covariance as a function of halo formation history and baryonic physics is a daunting task due to the highly non-linear and multi-scale physics involved in cluster formation. To make progress, in this paper, we adopt a simulation-based, data-driven approach whereby we develop semi-analytical parametric models of property covariance, which we then calibrate with cosmological simulations. We then apply our model to quantify the bias induced due to a non-zero property covariance in the expected weak lensing signal and the mass-observable scaling relation. 

As will be presented in \S\ref{sec:measurements}, a key element of this analysis is the estimation of true cluster \textit{richness} by encircling clusters within a 3D radius within the physical vicinity of the halo center, as opposed to a 2D projected radius used by cluster finders as redMaPPer \citep{Rykoff_2014} by identifying galaxies within the red-sequence band in the color-magnitude space --- the major difference being the removal of projection effects, or the mis-identification of non-cluster galaxies in the 2D projected radius from the photometric redshift uncertainty of the red-sequence when estimating the \textit{true} richness from a gravitationally bound region around the halo. Furthermore, as this simulation-based study does not introduce other observational systematics as shape noise of galaxies, point spread function, miscentering, among others, this study can be used to explore the \textit{intrinsic} covariance between observables prior to the addition of \textit{extrinsic} systematics as projection effects. Our results will not only provide insight into the physical origin of the covariance, the difference between the \textit{total} covariance as measured by observations and the \textit{intrinsic} covariance will provide estimates on the amplitude of the \textit{extrinsic} component.

The goals of this work are to (i) develop an analytical model that accounts for and quantifies the effect of non-zero covariance on cluster mass calibration, (ii) quantify this property covariance utilizing cosmological simulations, (iii) update uncertainties on inferred cluster mass estimates, and (iv) explain the physical origin of the covariance using secondary halo parameters. The rest of this paper is organized as follows. In \S\ref{sec:theory}, we present a population-based analytical framework. In \S\ref{sec:measurements}, we describe the simulations and data-vector employed in this work. In \S\ref{sec:cov_form}, we present our measurements and the covariance model. In \S\ref{sec:mass_calibration}, we present the impact of the covariance on weak lensing mass calibration. In \S\ref{sec:Xparams_cov}, we quantify the physical origin of the covariance by parameterizing it using secondary halo parameters. In \S\ref{sec:discussion}, we compare our work with those that employ realistic cluster finders. We conclude in \S\ref{sec:summary}. \\

\section{Theoretical Framework} \label{sec:theory} 

This section presents a theoretical framework that examines the impact of covariance on mass-observable scaling relations. In \S\ref{theory:obs} we introduce the definitions of richness and weak lensing excess surface mass density and their scaling relations with cluster mass. We then describe the model of property covariance of richness and excess surface mass density in \S\ref{sec:prob_statement}. In \S\ref{subsec{theory:wl_cov}}, we model the impact of covariance on stacked observable scaling relations. Finally, in \S\ref{subsec:Cov_Xparam_theory}, we develop a theoretical framework that explains the covariance based on a set of secondary halo parameters. A graphic representation of the outline of the paper is shown in Figure~\ref{fig:Outline}. The notations used in this section to describe the covariance are listed in Table \ref{tab:notation} and notations for scaling relations are listed in Table \ref{tab:scaling_notation}.

\begin{table}
\caption{Notations employed in our framework for the covariance in \S\ref{subsec{theory:wl_cov}}} \label{tab:notation}
	\begin{center}
		\tabcolsep=0.8mm
		\begin{tabular}{ | l | l | }
            \toprule
            \textbf{Parameter} & \textbf{Explanation}  \\ \midrule 
            $\Delta\Sigma$ & Weak lensing signal \\ 
            $M$ & halo mass in $M_{\odot}h^{-1}$ \\ 
            $N_{\rm gal}$ & optical richness enclosed inside 3D radius \\
            $z$ & redshift \\
            $r_{p}$ & projected and normalised radius \\
            \bottomrule
    \end{tabular}
	\end{center}
\end{table}

\begin{table}
\caption{Scaling Relation Conventions.} \label{tab:scaling_notation}
	\begin{center}
		\tabcolsep=0.8mm
		\begin{tabular}{ | l | l | }
            \toprule
            \textbf{Parameter} & \textbf{Explanation}  \\ \midrule 
            $\pi_{a}$ & normalization in scaling relation $\langle a \mid M\rangle$ \\
            $\alpha_{a}$ & slope in scaling relation $\langle a \mid M\rangle$ \\
            $\sigma_{a}$ & scatter about $\langle a \mid M\rangle$ \\
            $r_{a, b}$ & correlation between $a$ and $b$ at fixed $M$ \\
            $\pi_{a \mid b}$ & normalisation in scaling relation $\langle a \mid b\rangle$ \\
            $\alpha_{a \mid b}$ & slope in scaling relation $\langle a \mid b\rangle$ \\
            $\sigma_{a \mid b}$ & scatter about $\langle a \mid b\rangle$ \\
            \midrule 
            $a,b$ & $a,b \in \{\DSig, \ln\Ngal\}$ \\
            \bottomrule
    \end{tabular}
	\end{center}
\end{table}

\subsection{Observable-Mass Relations}
\label{theory:obs}

\subsubsection{Excess Surface Mass Density $\DSig$ from Weak Lensing} 
\label{theory:DeltaSigma}

In weak lensing measurements of galaxy clusters, the key observable is the excess surface mass density, denoted $\Delta \Sigma$. The excess surface mass density is defined as 
\begin{equation}
    \DSig(M, z, r_p) = \overline{\Sigma}(M, z, < r_p) - \Sigma(M, z, r_p),
\label{eqn:DeltaSigma}
\end{equation}
where $\overline{\Sigma}(M, z, < r_p) $ denotes the average surface mass density within projected radius $r_p$, and $\Sigma(M, z, r_p)$ represents the average of the surface mass density at $r_p$.
We model the average surface mass density $\Sigma$ as
\beqa
\Sigma(r_p) = \rho_m\int^{+\infty}_{-\infty}\Bigg(1+\xi_{hm}\Big(r = \sqrt{r_p^2 + \chi^2}\Big)\Bigg) d\chi,
\label{eqn:halo_matter_corr}
\eeqa
where $\rho_m$ is the mean matter density at the redshift of the cluster, $R$ is the projected radius in the plane of the sky, $\chi$ is the comoving distance along the line of sight centered around the cluster, and $\xi_{hm}(r)$ is the halo-matter correlation function which characterises the total mass density within a halo. 
Under the halo model, the halo-matter correlation function consists of a ``one-halo'' term:
\beqa
\xi_{\rm 1h} (r|M) = \frac{\rho_{\rm NFW}(r|M)}{\rho_{m0}} - 1,
\label{eqn:xi_NFW}
\eeqa
and a ``two-halo" term:
\beqa
\xi_{\rm 2h} (r|M) = b(M) \xi_{\rm lin}(r),
\label{eqn:xi_2h}
\eeqa
where $\rho_{\rm NFW}$ is the Navarro-Frenk-White (NFW) density profile \citep{NFW_1997}, and $\xi_{\rm lin}$ is the linear matter correlation function, and $b$ is the halo bias parameter. 

In weak lensing, the excess surface density $\Delta\Sigma$ is tied to the tangential shear $\gamma_t$ of the galaxies relative to the center of each foreground halo by the relation
\beqa
\Sigma_{\rm crit} \gamma_t = \overline{\Sigma}(<R) - \Sigma(R) \equiv \Delta\Sigma(R),
\label{eqn:DeltaSigma}
\eeqa
where the critical surface density $\Sigma_{\rm crit}$ defined as
\beqa
\Delta\Sigma_{\rm crit} = \frac{c^2}{4\pi G}\frac{D_s}{D_l D_{ls}},
\eeqa
and where $D_s$, $D_l$, and $D_{ls}$ refer to the angular diameter distances to the source, the lens, and between the lens and source, respectively. 

In this work, for each halo of mass $M$ at redshift $z$,  we compute the corresponding $\DSig$ profile. We compare these measurements with theoretical predictions --- in the one-halo regime we model the cluster overdensity as NFW profiles with their concentration determined by concentration-mass models of \citet{Prada_2012}, \citet{Ludlow_2016} and \citet{Diemer_19}, whereas in the two-halo term we adopt the linear matter correlation $\xi_{mm}$ multiplied by halo biases using the \citet{Tinker_2010}, \cite{Pillepich_2010} and \citet{Bhattacharya_2011} models to derive the halo-matter correlation $\xi_{hm}$. At the transition radius between the one- and two-halo regimes, we follow SDSS \citep{Zu_2014} in setting the halo--matter correlation to the maximum value of the two terms, i.e.,
\beqa
 \xi_{hm}(r|M) = \max\{\xi_{\rm 1h}(r|M),~\xi_{\rm 2h}(r|M) \}.
 \label{eqn:xi_max}
\eeqa
In Fig. \ref{fig:DS_200c_theory_data}, the theoretical models described above are compared with our measurements of $\Delta\Sigma$ in cosmological simulations to validate our data product. 

We model the mean $\langle \DSig \mid M, z, r_p \rangle$ at fixed mass $M$, redshift $z$, and projected radius $r_p$ as a log-linear relation given by
\begin{equation}
\langle \DSig \mid M, z, r_p \rangle_1 = \pi_{\DSig}(M, r_p, z) + \alpha_{\DSig}(M, r_p, z) \ln M, 
\label{eqn:model_DSig}
\end{equation} 
where $\alpha_{\DSig}$ is the power-law slope of the relation and $\pi_{\DSig}$ is a normalization that is a function of redshift and mass. 

\subsubsection{Optical Richness $\Ngal$} 
\label{theory:Ngal}

Optical richness $N_{\rm gal}$ is an observable measure of the abundance of galaxies within a galaxy cluster. It is often defined as the number of detected member galaxies brighter than a certain luminosity threshold within a given aperture or radius around the cluster centre. Richness is often used as a proxy for cluster mass, as more massive clusters are expected to have more member galaxies \citep[e.g.,][]{Rozo2014, Rykoff_2014}. The richness-mass scaling relation relates the richness of a galaxy cluster to its mass. In this work, we consider the mean $N_{\rm gal}$-$M_{\Delta}$ scaling relation expressed as
\begin{equation} 
    \langle \ln\Ngal  \mid M_{\Delta}, z \rangle_{1} = \pi_{\Ngal}(M, z) + \alpha_{\Ngal}(M, z) \ln M_{\Delta}.
    \label{eq:model_richness}
\end{equation}
where $M_{\Delta}$ is the mass of the halo within a radius where the mean density is $\Delta$ times the critical density of the universe, $\alpha_{\Ngal}(M, z)$ is the power-law slope of the relation, $\pi_{\Ngal}(M, z)$ is a normalisation that is a function of redshift and mass.

\subsubsection{Halo Mass and Radius Definitions}
\label{subsec{theory:halomass_radius}}

A common approach to defining a radial boundary of a galaxy cluster is such that the average matter density inside a given radius is the product of a reference overdensity $\Delta_{\rm ref}$ times the critical ($\rho_c$) or mean density ($\rho_{m}$) of the universe at that redshift. The critical density is defined as
\begin{equation}
    \rho_{c} = \frac{3H_0^2}{8\pi G}E(z),
\end{equation}
where $E(z)^2 = \Omega_{m,0}(1+z)^3 + \Omega_{\Lambda,0}$, $\Omega_{m,0}$ is the present day matter fraction of the universe, $\Omega_{\Lambda,0}$ is the dark energy fraction at the present age such that $\Omega_{m,0} + \Omega_{\Lambda,0} = 1$ for a flat universe ignoring the minimal contribution from the radiation fraction. The mean (background) density is defined as 
\begin{equation}
    \rho_{b} = \frac{3H_0^2}{8\pi G}(\Omega_{m,0}(1+z)^3).
\end{equation}

The overdensity $\Delta_c = 200$ is commonly chosen as the reference overdensity in optical weak lensing studies and is closely related to the virial radius. 
Another radius definition is the virial radius $R_{\rm vir}$, with overdensity values calibrated from cosmological simulations \citep{Bryan_Norman_1998}.
In this work, we use $R_{\rm 200c}$ and $R_{\rm vir}$ to scale various observations, including the $\Delta\Sigma$ measurements and richness. Since the covariance is close to zero at the outskirts $R \gtrsim R_{\rm 200c}$ as shown in \S\ref{sec:cov_form}, we adopt $r_p = R/R_{\rm 200c}$ and $r_p = R/R_{\rm vir}$ as our normalised radii, as the cluster properties are more self-similar with respect to $\rho_c(z)$ compared to $\rho_b(z)$ \citep{Diemer_2014, Lau_2015}. To test for the robustness of our covariance against different radii definitions, we also introduce a physical radius of a toy model of a constant $R = 1$~Mpc/h; here $h=0.6777$ is the reduced Hubble constant used in this study. 

\subsection{Covariance between $\DSig$ and $\Ngal$} \label{sec:prob_statement}

In optical surveys, we cannot expect the {\em covariance} between richness $N_{\rm gal}$ and the excess surface mass density $\DSig$ to be zero. Ignoring this covariance will lead to bias in cluster mass inferred from the excess surface mass density of the cluster selected based on richness. 
This work aims to quantify and analyse this covariance and its impact on the mass calibration relation. To achieve this objective, we must first specify the joint probability distribution of excess surface mass density and richness, $p(\DSig, \ln \Ngal \mid M, z, r_p)$. In this work, we assume that the joint distribution follows a multivariate normal distribution \citep{Stanek_2010, Evrard:2014, Mulroy2019,Miyatake2022}, which is fully specified with two components, the mean vector and the property covariance. We have checked the goodness of this assumption in Appendix~\ref{sec:cov_robustness}.

From the mean observable-mass scaling relations in Equation \eqref{eqn:model_DSig} and Equation \eqref{eq:model_richness}, the scaling relation between these two observables can be modeled as a local linear relation given by 
\begin{equation}
    \langle \DSig \mid \Ngal, z, r_p \rangle = \pi_{\DSig \mid \Ngal}(\Ngal, z, r_p) + \alpha_{\DSig \mid \Ngal}(\Ngal, z, r_p) \ln \Ngal,   
    \label{eq:Dsig_condition_Ngal}
\end{equation}
where $\pi$ and $\alpha$ are the normalization and slope of the model. 

The property covariance matrix is a combination of scatter and correlation between the scatter of $\DSig$ and $\ln N_{\rm gal}$ at a fixed halo mass, redshift, and projected radius. We use $\sigma_{\Ngal}(M, z)$ and $\sigma_{\DSig}(M, z, r_p)$ to denote the scatter of the observable-mass relation for $\ln N_{\rm gal}$ and $\DSig$, respectively, and use $r_{\Ngal, \DSig}(M, z, r_p)$ to denote the correlation between these scatters. The covariance matrix is then given by
\begin{equation}
    {\rm Cov}_{i,j}(M, z, r_p) = r_{i, j}(M, z, r_p)\ \sigma_{i}(M, z, r_p) \ \sigma_{j}(M, z, r_p),
\end{equation}
where $i$ and $j \in \{\DSig, \ln N_{\rm gal} \}$. 
Specifically, the covariance between $\DSig$ and $\Ngal$ can be expressed in terms of the residuals about the mean quantities
\begin{equation}
{\rm Cov}_{\DSig,\Ngal}(M, z, r_p) = {\rm Cov}({\rm res}_{\DSig}(M, z, r_p), {\rm res}_{\Ngal}(M, z)),
\end{equation}
where the residuals of 
the $\DSig$ and $\Ngal$ are, respectively, are given by
\begin{eqnarray}
{\rm res}_{\DSig}(M, z, r_p) = \DSig - \langle \DSig \mid M, z, r_p \rangle,  \\
{\rm res}_{\Ngal}(M, z) = \ln\Ngal- \langle \ln\Ngal \mid M, z \rangle.
\end{eqnarray}

To model the mass dependencies of the mean profiles of $\DSig$ and $\ln \Ngal$, we employ the Kernel Localised Linear Regression \citep[KLLR,][]{Farahi:2022KLLR} method. This regression method fits a locally linear model while capturing globally non-linear trends in data and has shown to be effective in modeling halo mass dependencies in scaling relations \citep[][]{Farahi2018,Wu_2022,Anbajagane:2022}. By developing a local-linear model of $\Delta\Sigma-\ln\Ngal$ with respect to the halo mass and computing the residuals about the mean relation, we remove the bias in the scatter due to the mass dependence and reduce the overall size of the scatter. As shown in Fig. \ref{fig:cov_function_compare} the $1-\sigma$ of the covariance is determined by bootstrap resampling. 

\subsection{Corrections to the $\DSig-\Ngal$ relation due to Covariance} \label{subsec{theory:wl_cov}}

The shape of the halo mass function plays an important role in evaluating the conditional mean value of $\langle \DSig \mid \Ngal, z, r_p \rangle$ where the scatter between two observables with a fixed halo mass is correlated. 
Ignoring the contribution from the correlated scatter, to the zeroth order, the expected $\DSig$ evaluated at fixed richness is given by Equation \eqref{eq:Dsig_condition_Ngal}. This is the model that has been used in mass calibration with stacked weak lensing profiles \citep[][]{Johnston_07, Kettula_15, McClintock2019, Chiu_20, Lesci_22}. 

The first- and second-order approximations of the scaling relation are given by
\begin{align} 
    \langle \DSig \mid \Ngal, z \rangle_1 &= \langle \DSig \mid \Ngal, z \rangle_{\rm fid} + \frac{\gamma_1}{\alpha_{\Ngal}} \times {\rm Cov}(\DSig, \ln\Ngal), \label{eq:1st_order_correction} 
\end{align}
and
\begin{align}  \label{eq:2nd_order_correction_s_b} 
    \langle \DSig  & | \Ngal,z \rangle_2 = \langle \DSig \mid \Ngal, z \rangle_{\rm fid} + \\ \nonumber
     &{\rm Cov} (\DSig, \ln \Ngal) \times 
     \Big[\frac{x_s}{\alpha^2_{\Ngal}} (\alpha_{\Ngal} \gamma_1 + \gamma_2(\ln \Ngal - \pi_{\Ngal}))  \Big], 
\end{align}
where $\langle \DSig \mid \Ngal, z \rangle_{\rm fid}$ is the fiducial relation taking into account the curvature of the HMF but independent of the covariance; $x_{s} = (1 + \gamma_2\sigma^2_{M|N_{\rm gal},1})^{-1}$ is the compression factor due to curvature of the HMF, the subscript 1 denoting that the scatter is taken from the HMF expanded to first order; here we omit the $(M,z)$ dependence of the covariance as a shorthand notation. These expansions around the pivot mass are for halos centered around a narrow enough mass bin. We show explicitly in Fig. \ref{fig:DS_cov_correction} that the first-order expansion converges using our binning method. The derivations for the first and second-order expansion terms can be found in \citet{Evrard:2014} and \citet{Farahi_2018} and the derivation for this particular expression of the second-order term is shown in Appendix \ref{sec:derive_2nd_order}.

Here $\gamma_1$ and $\gamma_2$ are the parameters for the first and second-order approximations to the mass dependence of the halo mass function \citep[e.g.,][]{Evrard:2014}:
\begin{equation}
\frac{{\rm d}n_{\rm hmf}(M, z)}{{\rm d}\ln M} \approx A(z) \exp\left[- \gamma_1(M, z) \ln M  -\frac{1}{2}\gamma_2(M, z)(\ln M)^2\right].  
\end{equation} 
where $A(z)$ is the normalisation of the mass function due to the redshift alone. 
In deriving the above approximations, we have made use of the fact that ${{\rm Cov}}(\DSig, \ln \Ngal | M, z) \equiv r_{\DSig, \Ngal}\sigma_{\DSig}\sigma_{\Ngal}$.
The terms $\sigma_{\DSig}$, $\sigma_{\Ngal}$, $r_{\Ngal, \DSig}$, $\gamma_1$ and $\gamma_2$ are evaluated at the mass implied by $\langle \ln M \mid N_{\rm gal} \rangle$.

These property covariance correction terms are absent in the current literature. A key feature of this approximation method is that the second-order solution has better than percent-level accuracy when the halo mass function is known \citep{Farahi_2016}. In Figure~\ref{fig:DS_cov_correction}, we demonstrate that the statistical uncertainties for the first-order correction in Equation~\eqref{eq:1st_order_correction} is larger than the uncertainty in the halo mass function and the uncertainty due to the second-order halo mass approximation. 

\subsection{Secondary halo parameter dependence of ${\rm Cov}(\DSig, \ln \Ngal | M, z)$}
\label{subsec:Cov_Xparam_theory}

\begin{table}
\caption{Notations employed in exploring the secondary halo parameter dependence} 
\label{tab:notation_Xparam}
	\begin{center}
		\tabcolsep=0.8mm
		\begin{tabular}{ | l | l | }
            \toprule
            \textbf{Parameter} & \textbf{Explanation}  \\ \midrule 
            $\Pi$ & Set of secondary halo parameters \\
            $\Gamma_{\rm inst}$ & instantaneous mass accretion rate (MAR) \\ 
            $\Gamma_{\rm 100Myr}$ & mean MAR over the past 100 Myr \\ 
            $\Gamma_{\rm dyn}$    & mean MAR over virial dynamical time \\ 
            $\Gamma_{\rm 2dyn}$   & mean MAR over two virial dynamical time \\ 
            $\Gamma_{\rm peak}$     & Growth rate of peak mass from current z to z+0.5 \\ 
            $a_{1/2}$               & Half mass scale factor \\ 
            $c_{\rm vir}$           & $R_{\rm vir}$ concentration \\ 
            T/|U|                   & Absolute value of the kinetic to potential energy ratio \\ 
            $X_{\rm off}$          & Offset of density peak from mean particle position (kpc~$h^{-1}$) \\
            \bottomrule
        \end{tabular}
	\end{center}
\end{table}

We elucidate the physical origin of the covariance between $\DSig$ and $\ln\Ngal$ by developing a phenomenological model based on the secondary halo parameters listed in Table~\ref{tab:notation_Xparam}. These parameters are computed from the ROCKSTAR halo finder \citep{Behroozi_2013}. They capture the halo's mass accretion history , which we hypothesise is the driving force behind the observed covariance. To incorporate these parameters into our model, we extend Equations~\eqref{eqn:model_DSig} and \eqref{eq:model_richness}  by introducing multi-linear terms that include the secondary halo parameters denoted by the vector $\Pi$:
\begin{align}
    ( \ln \Ngal \mid \Pi, M, z ) =& \langle \ln\Ngal \mid M, z \rangle_1 + \vec{\beta}_{\Ngal}^\intercal(M, z) \cdot \Pi + \epsilon_{\Ngal} \nonumber \\
    =& \pi_{\Ngal}(M, z) + \alpha_{\Ngal}(M, z) \ln M + \nonumber\\
     &\vec{\beta}_{\Ngal}^\intercal(M, z) \cdot \Pi + \epsilon_{\Ngal}. \label{eqn:model_richness_Xparam} \\
    ( \DSig \mid \Pi, M, z )  =& \langle \DSig \mid M, z \rangle_1 + \vec{\beta}_{\DSig}^\intercal(M, z) \cdot \Pi + \epsilon_{\DSig} \nonumber \\
    =& \pi_{\DSig}(M, z) + \alpha_{\DSig}(M, z) \ln M + \nonumber  \\ 
    & \vec{\beta}_{\DSig}^\intercal (M, z) \cdot \Pi + \epsilon_{\DSig}, \label{eqn:model_wl_Xparam}
\end{align}
where $\Pi$ is a vector of secondary halo parameters of potential interest listed in Table~\ref{tab:notation_Xparam}, and $\epsilon_{\DSig}$ and $\epsilon_{\Ngal}$ are normally distributed intrinsic scatter terms with zero means and uncorrelated variances. In Appendix \ref{sec:multilinear_modeling} we show that the residual conditioned on secondary halo parameters can largely be assumed to be Gaussian. Additionally, we assume $\langle \epsilon_{\Ngal} \epsilon_{\DSig} \rangle = 0$, which implies that the scatter about the mean relations is uncorrelated after factoring in the secondary properties.

Due to the bilinearity and distributive properties of covariance, combining  Equation~\eqref{eqn:model_richness_Xparam} and Equation~\eqref{eqn:model_wl_Xparam} yields:
\begin{align} \label{eq:cov_Xparam_model}
    \text{Cov}(\DSig, & \ln\Ngal \mid M,z) = \text{Cov}(\langle \DSig \rangle_1, \langle \Ngal \rangle_1) + \text{Cov}(\langle  \DSig \rangle_1, \vec{\beta}_{\Ngal}^\intercal \cdot \Pi) \nonumber \\ 
    & \ \  + \text{Cov}(\langle  \DSig \rangle_1, \epsilon_{\Ngal}) + \text{Cov}(\vec{\beta}_{\DSig}^\intercal \cdot \Pi, \langle \ln\Ngal \rangle_1) \nonumber \\ 
    & \ \  + \text{Cov}(\vec{\beta}_{\DSig}^\intercal \cdot \Pi, \vec{\beta}_{\Ngal}^\intercal \cdot \Pi) + \text{Cov}(\vec{\beta}_{\DSig}^\intercal \cdot \Pi, \epsilon_{\Ngal}) \nonumber \\ 
    & \ \   + \text{Cov}(\epsilon_{\DSig}, \langle \ln\Ngal \rangle_1) + \text{Cov}(\epsilon_{\DSig}, \vec{\beta}_{\Ngal}^\intercal \cdot \Pi) \nonumber \\
    & \ \  + \text{Cov}(\epsilon_{\Ngal}, \epsilon_{\DSig}),
\end{align}
where we omit the explicit $(M,z)$ dependence in $\langle \DSig \mid M,z \rangle_1$, $\langle \ln\Ngal \mid M,z \rangle_1$, $\vec{\beta}_{\DSig}^\intercal(M, z)$, $\vec{\beta}_{\Ngal}^\intercal(M,z)$, $\epsilon_{\Ngal}(M,z)$ and $\epsilon_{\DSig}(M,z)$ to simplify the notation. The KLLR method is utilised to estimate these mass-dependent parameters. All terms involving $\langle \DSig\rangle$ and $\langle \ln\Ngal \rangle$ vanish, as these terms are independent of $\Pi$ by definition. Terms involving $\epsilon_{\Ngal}$ and $\epsilon_{\DSig}$ also go to zero, as they are uncorrelated Gaussian scatters.  Only the term ${\rm Cov}(\vec{\beta}_{\DSig}^\intercal \cdot \Pi, \vec{\beta}_{\Ngal}^\intercal \cdot \Pi)$ remains, and hence our final expression for the covariance is
\begin{align} \label{eq:cov_Xparam_model_part2}
    {\rm Cov}(\DSig,\ln\Ngal \mid M,z) &= {\rm Cov}(\vec{\beta}_{\DSig}^\intercal \cdot \Pi, \vec{\beta}_{\Ngal}^\intercal \cdot \Pi) \\ \nonumber
    & = \vec{\beta}_{\DSig}^\intercal {\rm Cov}(\Pi,\Pi) \vec{\beta_{\Ngal}}.
\end{align}

To estimate the error in the covariance due to each of the secondary halo parameters, we compute ${\rm Cov}(\Delta\Sigma, \Pi_i| M,z)$ for each secondary halo parameter $i$ in each ($r_p$, $M$, $z$) bin and take their standard deviations as the error measurement. Modeling the richness-mass relation as in Equation \eqref{eqn:model_richness_Xparam} and using the same derivation as in Equation~\eqref{eq:cov_Xparam_model}, we arrive at the expression
\begin{equation}
    {\rm Cov}(\Delta\Sigma, \ln\Ngal | M, z) = \sum_{i}\beta_{\Ngal,i}(M,z)\ {\rm Cov}(\Delta\Sigma, \Pi_i| M, z),
    \label{eqn:cov_hypothesis}
\end{equation}
in which the error from each contributing term in $\Pi$ is the standard deviation for ${\rm Cov}(\Delta\Sigma, \Pi_i| M, z)$ multiplied by the partial richness slope $\beta_{\rm \Ngal, i}$. The total variance of ${\rm Cov}(\Delta\Sigma, \ln\Ngal | M, z)$ are the errors of each term added in quadrature.

We test the validity of this model by checking how well the secondary halo parameters can explain covariance between lensing and richness in \S\ref{sec:Xparams_cov}. After subtracting the covariance from each of the $\Pi_i$ terms, the full covariance should be consistent with null, given the uncertainty. Our results confirm that the dependency of secondary halo parameters can indeed explain the covariance.

\section{Dataset and measurements}\label{sec:measurements}

In this section, we describe the measurements on the individual ingredients that make up the covariance --- $\Delta\Sigma$ the lensing signal in \S\ref{sec:DS_model} and $\ln{N_{\rm gal}}$ the log-richness measurement in \S\ref{sec:Ngal_model}.

\subsection{Measurements of $\Delta\Sigma$} \label{sec:DS_model}

We employ the MultiDark Planck 2 (MDPL2) cosmological simulation \citep{Klypin2016} to measure halo properties. The MDPL2 is a gravity-only $N$-body simulation, consisting of $3840^3$ particles in a periodic box with a side length of $L_{\rm box}=1~h^{-1}{\rm Gpc}$, yielding a particle mass resolution of approximately $m_{\rm p} \approx 1.51 \times 10^9 h^{-1}M_\odot$. The simulation was conducted with a flat $\Lambda$CDM cosmology similar to \citet{Planck2014}, with the following parameters: $h=0.6777$, $\Omega_{\rm m} = 0.307115$, $\Omega_\Lambda = 0.692885$, $\sigma_8 = 0.829$, and $n_{\rm s} = 0.96$. We use the surface over-density of down-sampled dark matter particles to measure the weak lensing signal. We selected cluster-sized halos using the ROCKSTAR \citep{Behroozi_2013} halo catalogue, which includes the primary halo property of mass and redshift and a set of secondary halo properties listed in Table \ref{tab:notation_Xparam} that we utilise in this analysis. To capture the contribution of both the one- and two-halo terms to $\xi_{hm}$, we use a projection depth of $D_p = 200$ Mpc $h^{-1}$ to calculate $\Delta\Sigma$ \citep[e.g.,][]{Costanzi_2019, Sunayama_2020}.  The MDPL2 data products are publicly available through the MultiDark Database \citep{Riebe2013} and can be downloaded from the {\tt CosmoSim} website\footnote{\url{https://www.cosmosim.org}}. 

\begin{figure*}
    \centering
    \includegraphics[width=\linewidth]{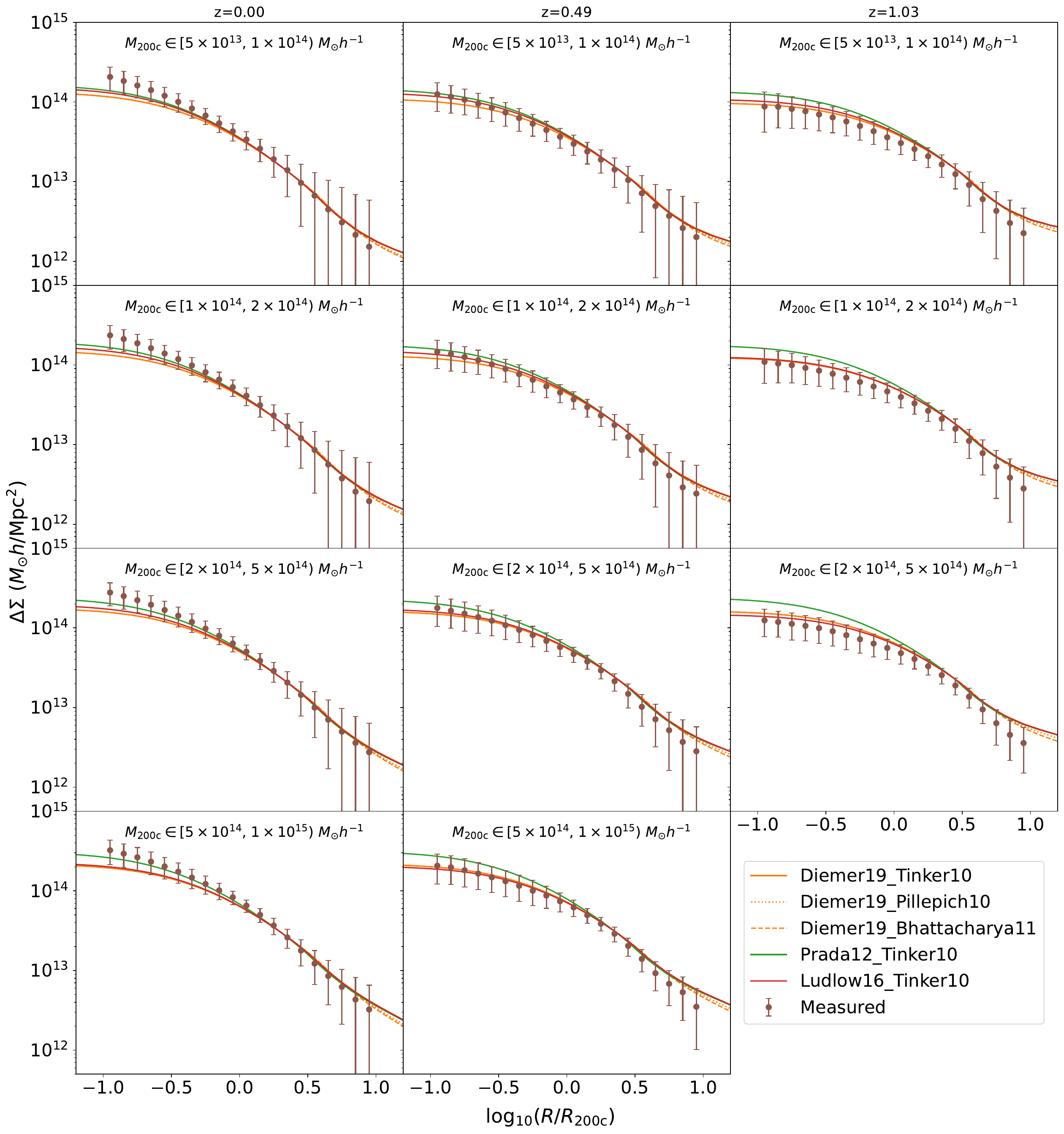}
    \caption{The measured $\Delta\Sigma$ profiles using downsampled particles for every 10 particles and theoretical $\Delta\Sigma$ as computed from the NFW profile using different concentration-mass relations (LHS in legend) in the one-halo regime and different halo-bias models (RHS in legend) in the two halo regimes, with errors taken to be $1-\sigma$ standard deviations; the measurements are consistent with theoretical predictions and the size of the errors is too large to distinguish between models. The same conclusion (not shown) holds for $\Delta\Sigma$ binned by $R_{\rm vir}$. }
    \label{fig:DS_200c_theory_data}
\end{figure*}

The excess overdensity, $\Delta\Sigma$, is calculated by integrating the masses of the dark matter particles in annuli of increasing radius centred around the halo centre. However, since clusters do not have a well-defined boundary, we compare the results of two radial binning schemes. The first scheme uses 20 equally log-spaced ratios between 0.1 and 10 times $R_{\rm vir}$, while the second scheme spans 0.1 to 10 times $R_{\rm 200c}$. We consider the measurements binned at $R_{\rm 200c}$ as our final results to be consistent with the weak lensing literature. Figure~\ref{fig:DS_200c_theory_data} shows that our measurements are consistent with most models of the concentration-mass and halo-bias models at a $1\sigma$ level. 

At a projection depth of $D_p = 200$ Mpc $h^{-1}$, the projection effects can be modeled as a multiplicative bias \citep{Sunayama_2023}. In \citet{Sunayama_2023} the projection effects on $\Delta\Sigma$ are modelled as $\Delta\Sigma_{\rm obs} = (1 + \alpha)\Delta\Sigma_{\rm true}$, where $\alpha = 18.4 \pm 8.6\%$. Although the multiplicative bias of projection effects may increase the amplitude of ${\rm Cov}(\Delta\Sigma, \ln N_{\rm gal})$ by a factor of $(1+\alpha)$, we argue that it does not introduce an additive bias into our model for ${\rm Cov}(\Delta\Sigma, \ln N_{\rm gal})$. This is because, under the richness model and $\Delta\Sigma$ in Equations~\eqref{eqn:model_richness_Xparam} and~\eqref{eqn:model_wl_Xparam}, only terms in richness that are correlated with projection effects will contribute to the covariance. As demonstrated in \S\ref{sec:Ngal_model}, we enclose the halo within a 3D physical radius, so the number count $N_{\rm gal}$ of galaxies should not include projection effects. Therefore, projection effects should not introduce an additive bias to our covariance.

To remove the 2D integrated background density, we first computed the background density of the universe ($\rho_b$) at the cluster redshift using the cosmological parameters of the MDPL2 simulation. The integrated 2-D background density is given by $\Sigma_b = 2D_p \rho_b$, where factor 2 comes from the integration of the foreground and background densities. 

\subsection{Measurements of $N_{\rm gal}$} \label{sec:Ngal_model}
\subsubsection{Dataset for $N_{\rm gal}$ --- SAGE galaxy catalog}

The Semi-Analytic GALAXY Evolution (SAGE) is a catalogue of galaxies within MDPL2, generated through a post-processing step that places galaxies onto N-body simulations. This approach, known as a semi-analytic model (SAM), is computationally efficient compared to hydrodynamical simulations with fully self-consistent baryonic physics. SAMs reduce the computational time required by two to three orders of magnitude, allowing us to populate the entire 1~(Gpc/h)$^3$ simulation volume with galaxies. SAGE's statistical power enables us to conduct stacked weak lensing analyses.

The baryonic prescription of SAGE is based on the work of \cite{Croton_2016}, which includes updated physics in baryonic transfer processes such as gas infall, cooling, heating, and reionization. It also includes an intra-cluster star component for central galaxies and addresses the orphan galaxy problem by adding the stellar mass of disrupted satellite galaxies as intra-"cluster" mass. SAGE's primary data constraint is the stellar mass function at $z = 0$. Secondary constraints include the star formation rate density history \citep{Somerville2001}, the Baryonic Tully-Fisher relation \citep{Stark2009}, the mass metallicity relation of galaxies \citep{Tremonti_2004}, and the black hole--bulge mass relation \citep{Stark2009}.

\subsubsection{Model for  $N_{\rm gal}$} 

\begin{figure}
    \centering
    \includegraphics[width=\linewidth]{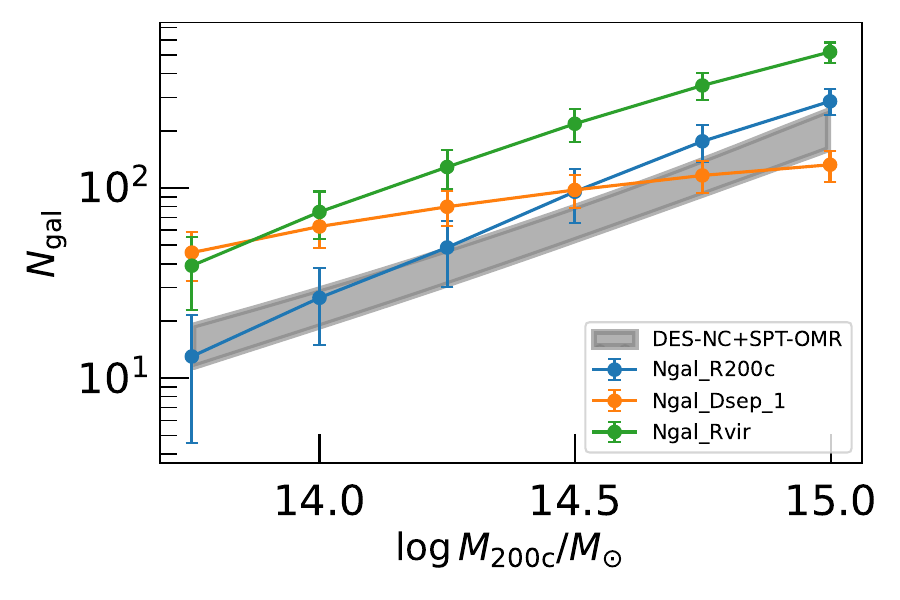}
    \caption{Using different prescriptions of the richness count, we compare with the SPT-DES data \citep{Costanzi_2021}. The richness estimator, with no stellar or color-magnitude cut, shows a similar trend with the data. In \S\ref{sec:cov_form}, we show that the results are robust to changes in the definition of the number count estimator.}
    \label{fig:MOR_DMPL2_vs_data}
\end{figure}

To determine the number of galaxies inside a cluster-sized halo, we utilise the SAGE semi-analytic model and compute the total number of galaxies within a 3D radius around the halo centre. We compare the true richness ($N_{\rm gal}$) to $M_{\rm 200c}$ scaling relations between different models and the observed richness-mass relations from \citet{Costanzi_2021} using data from the DES Year-1 catalogue and mass-observable-relation from the South Pole Telescope (SPT) cluster catalogue (see Figure~\ref{fig:MOR_DMPL2_vs_data}). The observed richness-mass relation is fitted as a log-linear model with $2-\sigma$ error bars that trace the posterior of the best-fit richness-mass model parameters. The $M_{\rm 500c}$ mass definition in the catalogue is converted to $M_{\rm 200c}$ using an NFW profile for the surface density of the cluster and adopting the \citet{Diemer_19} concentration-mass relation anchored at $z=0.35$, which is roughly the median redshift of the cluster sample.

We use the KLLR method to determine the local linear fit for our $N_{\rm gal}$-mass model, which relaxes the assumption of global log-linearity \citep[][]{Anbajagane:2020stellar}. Realistic cluster finders, such as redMaPPer \citep{Rykoff_2014}, impose a colour-magnitude cut or a stellar mass cut, which are highly dependent on the red-sequence model or the spectral energy density model. We found that imposing a stellar mass cut of $10.5\log(M/M_{\odot})$ would correspond roughly to the bottom 5\% percentile of SDSS detected galaxies \citep{Maraston_2013}. However, this drastically decreases the number of galaxies in a halo, with most having $N_{\rm gal}$ in the single digits. As we are interested in the \textit{intrinsic} covariance from the physical properties of the halo, we do not impose additional magnitude or stellar mass cuts. We confirm that, as described in \citet{Croton_2016}, the galaxy stellar mass distribution at $z=0$ is consistent with the best-fit double Schechter function calibrated with low-redshift galaxies from the Galaxy and Mass Assembly (GAMA) \citep{Baldry_2012} down to stellar masses of $\mathcal{M} > 10^{8.5} M_{\odot}$.

Figure~\ref{fig:MOR_DMPL2_vs_data} illustrates that our $N_{\rm gal}$-mass models, which count the number of galaxies within a physical 3D radius and impose no colour-magnitude cut as redMaPPer does, resemble the general behavior of the observed richness-mass relations in terms of both slope and intercept. However, we acknowledge that redMaPPer may suffer from projection effects that artificially inflate the number count of red-sequence galaxies within its aperture because of line-of-sight structures. Additionally, the $N_{\rm gal}$ count within the $R_{\rm vir}$ radius exceeds that of $R_{\rm 200c}$ as $R_{\rm vir}$ is greater than $R_{\rm 200c}$. In the toy model scenario, where we use a constant $1$ Mpc $h^{-1}$ radius, the slope of the mass-richness relation starts to decrease as the mass increases due to the increasing physical size of the clusters, as expected. The diversity of cluster radii and the resulting variation in the local slope and intercept of the $N_{\rm gal}$-mass relations demonstrate the robustness of our covariance model. In Section \ref{sec:cov_form}, we show that different radii/mass definitions have little impact on the parameters of our covariance model, thus establishing its independence from different reference radii, the definitions of cluster edges, and the resulting richness-mass relations.

\section{Results: Covariance Shape and Evolution}
\label{sec:cov_form}

\begin{figure*}
    \centering
    \includegraphics[width=0.9\textwidth]{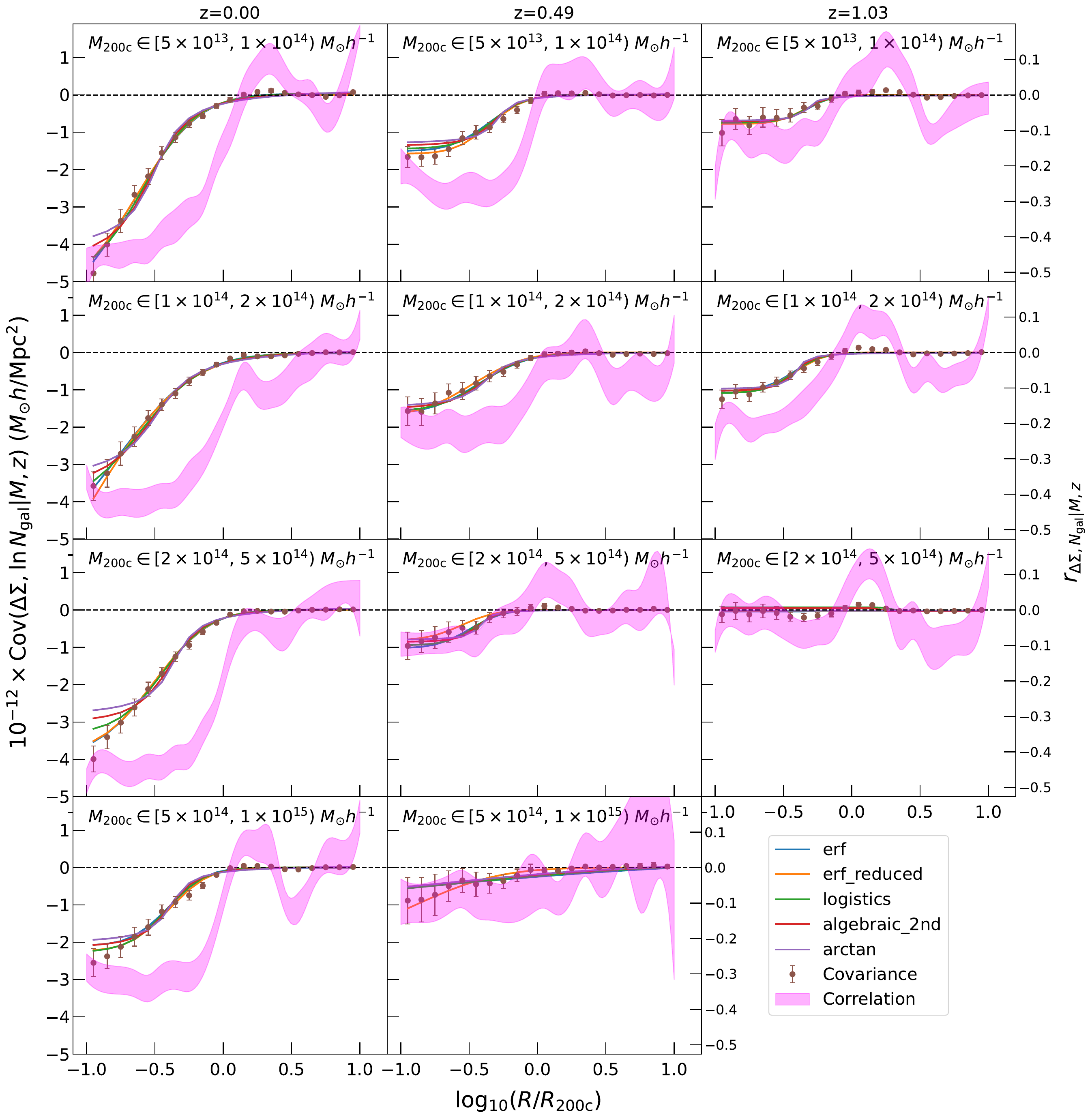}
    \caption{Measured against the left hand side y-axis are measurements of Cov($\Delta\Sigma$, $N_{\rm gal}|M,z$) with $1\sigma$ errors and different functional forms using the full model. The functions are classes of "Sigmoid" functions. In all bins, the error function outperforms other functional forms in their DIC parameters, providing good $\chi^2$ values. For $M_{\rm 200c} \in [5\times10^{14},1\times10^{15})$ at $z=0.49$ and $M_{\rm 200c} \in [2\times10^{14},5\times10^{15})$ at $z=1.03$, the posteriors of the full models do not converge as the size of the covariance is too small. Measured against the right-hand side y-axis are the correlation coefficients $r_{\DSig,\Ngal | M,z}$ with smoothed bands representing the $1-\sigma$ error. The errors are measured by bootstrap resampling.}
    \label{fig:cov_function_compare}
\end{figure*}

In this section, we report the measurements for our covariance. In Figure~\ref{fig:cov_function_compare}, we find an anti-correlation between $N_{\rm gal}$ and $\Delta\Sigma$ at small scales across most redshift and mass bins spanned by our dataset, which we fit with the best-fit ``Sigmoid'' functional form of the expression 
\beqa
\rm{Cov}(\tilde{x}) = s\Big( {\rm erf}(\frac{\sqrt{\pi}}{2} \tilde{x}) + g\Big),
\eeqa
with $x \equiv \log{R/R_{\rm 200c}}$ the log-radius and $\tilde{x} \equiv (x-\gamma)/\tau$ the scaled and offset log-radius. In Appendix \ref{sec:functional_form}, we offer statistical verification of the best-fit functional form.

We first describe the evolution of the covariance in \S\ref{subsec:cov_binned_mass} by binning across the $(M,z)$ bins. Next, in \S\ref{subsec:cov_binned_peakheight}, we present an alternative binning scheme based on halo peak height that can provide insight into the dependence of the time formation history of the covariance scale.

\subsection{Binned in $(M,z)$}
\label{subsec:cov_binned_mass}

Our best-fit parameters in Table \ref{tab:cov_best_fit_param_full_model} indicate that in 9 out of 12 $(M,z)$ bins, the Cov($\Delta\Sigma, \ln N_{\rm gal} \mid M, z)$  rejects the null-correlation hypothesis with high statistical significance ($p-{\rm value} < 0.01$). However, in two bins, specifically $M_{\rm 200c} \in [5\times10^{14},1\times10^{15})\ M_{\odot}h^{-1}$ at $z=0.49$ and $M_{\rm 200c} \in [2\times10^{14},5\times10^{15})\ M_{\odot}h^{-1}$ at $z=1.03$, the magnitude of the covariance is relatively small compared to the size of their errors. Consequently, it becomes challenging to constrain the shape parameters in these two bins, and the covariance is consistent with the null hypothesis. Furthermore, we exclude the bin $M_{\rm 200c} \in [5\times10^{14},1\times10^{15})\ M_{\odot}h^{-1}$, $z=1.03$ due to the limited number of halos it contains.

Our results suggest that the shape of the covariances can be accurately described by the full error function. Additionally, for $R \geq R_{\rm vir}$ or $R \geq R_{\rm 200c}$, the covariance aligns with the null-correlation hypothesis. This alignment is reflected in the fact that all nine bins with constrained posterior shape have best-fit $g$ values within $2\sigma$ of $g=-1$. Deviations from $g=-1$ can be interpreted as evidence of disagreements with the Press-Schechter formalism \citep{Press_Schechter_1974} of spherical collapse halos, which can be originated from the presence of anisotropic or non-Gaussian matter distribution around halos at large scales \citep{Lokken_2022}, or it can be an indicator of an open-shell model of halos that allows for the bulk transfer of baryonic and dark matter in and out of the halo potential well during the non-linear collapse.

With $g=-1$ fixed, the reduced error function marginally improves the constraints in most bins. However, with the reduced model, we can provide posterior constraints for $M_{\rm 200c} \in [5\times10^{14},1\times10^{15})M_{\odot}h^{-1}$ at $z=0.49$ and $M_{\rm 200c} \in [2\times10^{14},5\times10^{15}) M_{\odot}h^{-1}$ at $z=1.03$, which the full model failed to constrain but with very loose posterior constraints. The estimated parameters for both the full and reduced models are presented in Table~\ref{tab:cov_best_fit_param_full_model} and Table~\ref{tab:error_func_reduced}, respectively.

\begin{figure}
    \centering
    \includegraphics[width=\linewidth]{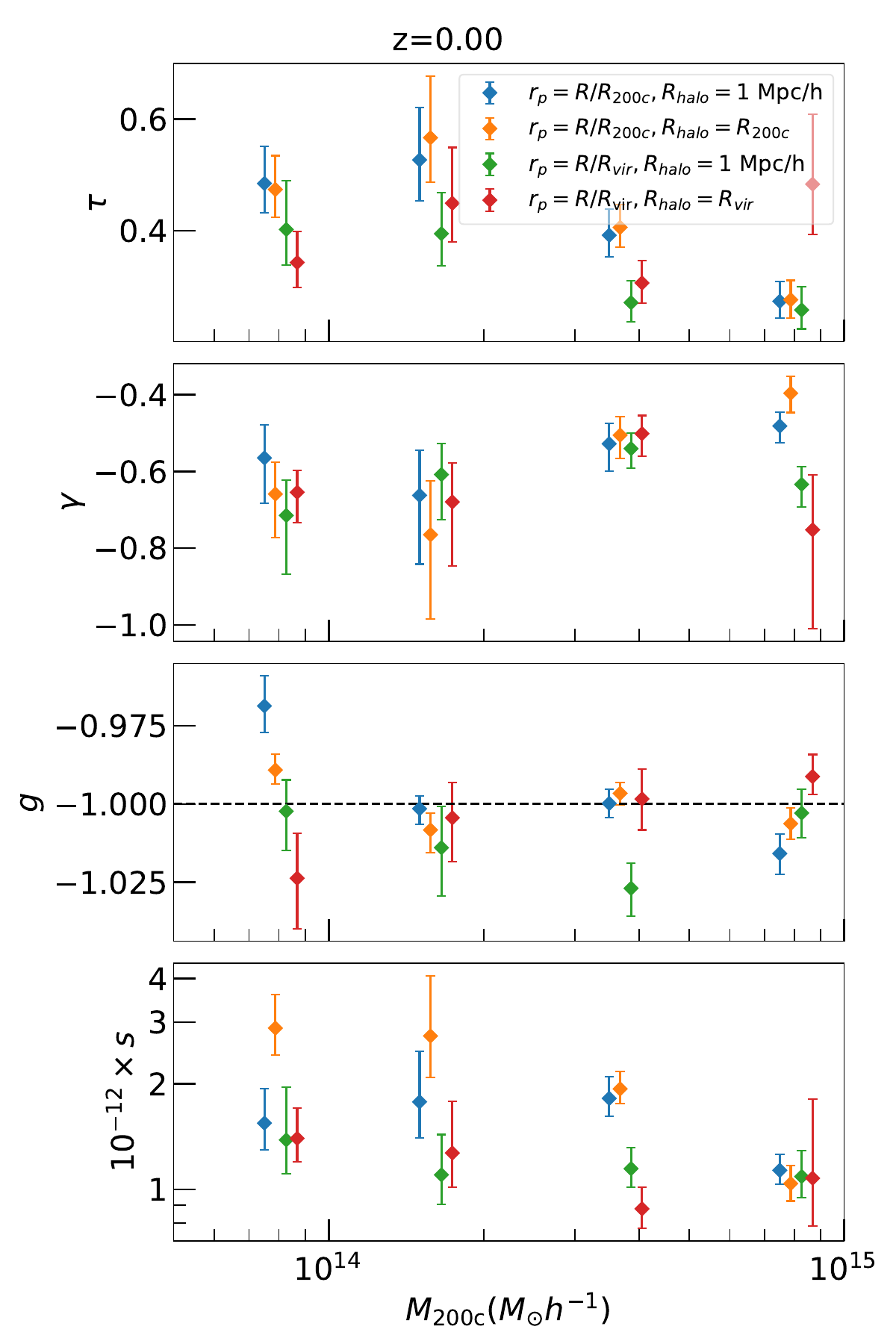}
    \caption{Evolution of Cov($\Delta\Sigma$, $N_{\rm gal} \mid M, z$) shape parameters with respect to mass and radial binning schemes and $N_{\rm gal}$ definition at fixed redshift at z=0. $\Delta\Sigma$ is binned in equal log-space radial bins in $R/R_{\rm 200c}$ or $R/R_{\rm vir}$; for each radial binning, the number count of galaxies inside the cluster is given by a constant radius of $1\ {\rm Mpc} h^{-1}$ or $R_{\rm 200c}$ when binned by $R_{\rm 200c}$ and $R_{\rm vir}$ when binned by $R_{\rm vir}$. We find no strong evolution in the shape or scale of the covariance under different binning schemes or $N_{\rm gal}$ definitions. The trend is consistent across different redshift bins and demonstrates the robustness of the covariance under different true richness definitions. The error bars indicate the $1-\sigma$ distribution of the posteriors. }
    \label{fig:cov_params_vary_Ngal}
\end{figure}

To assess the impact of varying the definition of the halo radius on our measurements of the shape of the covariance, we considered two factors: the scale dependence of $\Delta\Sigma$ discussed in \S\ref{subsec{theory:halomass_radius}} and \S\ref{sec:DS_model}, and the alteration of the richness-mass relation as shown in Figure \ref{fig:MOR_DMPL2_vs_data} in \S\ref{sec:Ngal_model}. Figure \ref{fig:cov_params_vary_Ngal} demonstrates that there is no apparent evolution of the shape parameters $\boldsymbol{\theta} \in \{\tau,\gamma, g\}$ when altering the scale dependence for $\Delta\Sigma$ or the true richness count. However, we find marginal $3\sigma$ evidence of a difference in the amplitude parameter of the covariance $s$ when changing the scale normalisation from $r_p = R/R_{\rm 200c}$ to $r_p = R/R_{\rm vir}$ while using the same true richness count. As halos exhibit more self-similarity in the inner regions when scaled by $R_{\rm 200c}$ \citep{Diemer_2014}, we adopt this as our radius normalisation and use the number of galaxies enclosed within $R_{\rm 200c}$ as our true richness count.

\begin{figure}
    \centering
    \includegraphics[width=\linewidth]{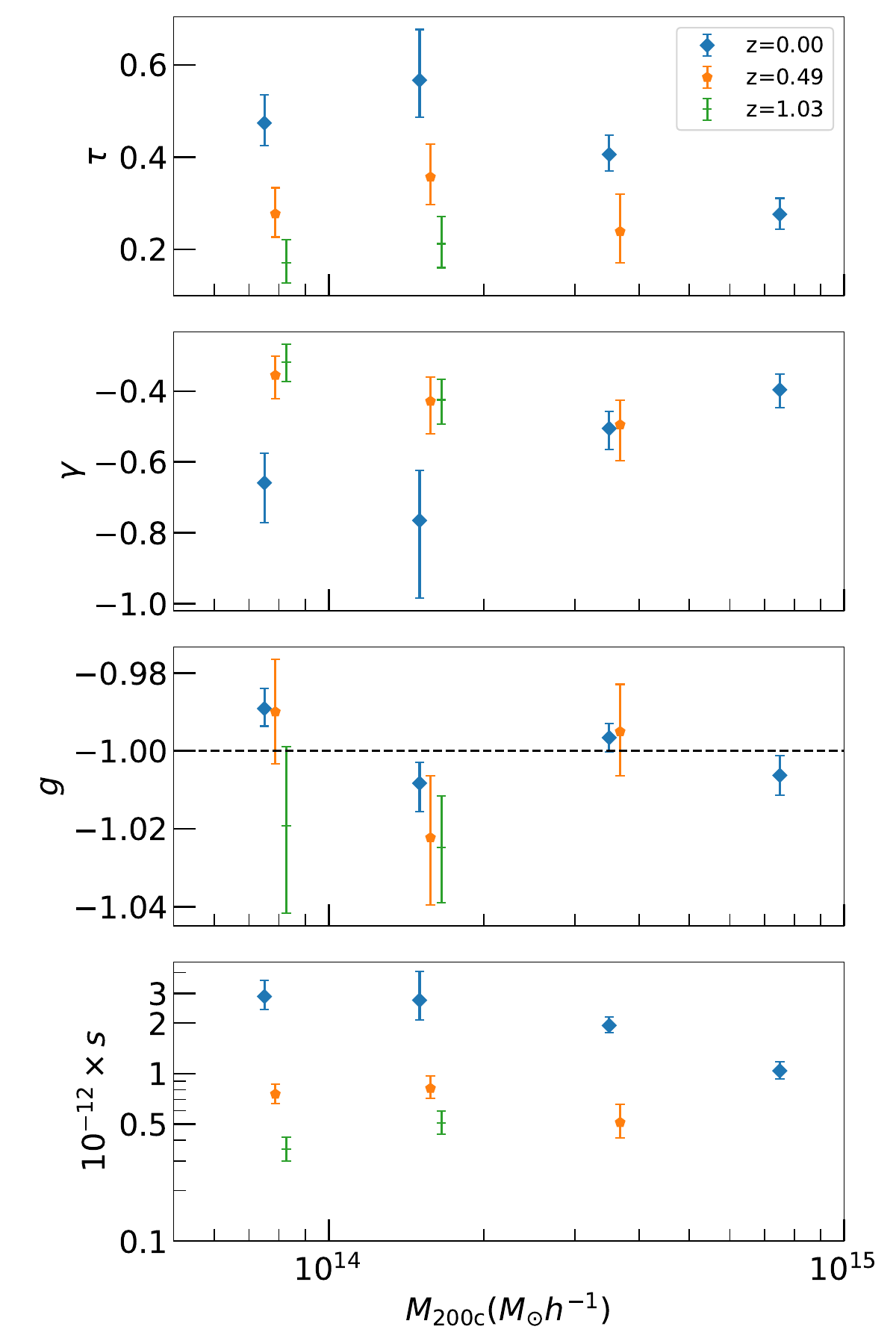}
    \caption{Evolution of Cov($\Delta\Sigma$, $N_{\rm gal} \mid M, z$) shape parameters of the error function with respect to mass and redshift, binned in units of $R_{\rm 200c}$ and with $N_{\rm gal}$ taken to be the number of clusters inside the $R_{\rm 200c}$ radius of the cluster. There is no strong dependence of $\tau$, $\gamma$, and $g$ with respect to mass and redshift and a strong monotonically decreasing $s$ with respect to redshift. At $M_{\rm 200c} \in [5\times10^{14},1\times10^{15})$, $z=0.49$ and $M_{\rm 200c} \in [2\times10^{14},5\times10^{15})$, $z=1.03$ the covariance is consistent with null at $p =0.01$ and $p=0.05$ levels, respectively. The error bars indicate the $1-\sigma$ distribution of the posteriors.
    }
    \label{fig:cov_params_vary_mass_redshift}
\end{figure}

Subsequently, we explored the evolution of the shape parameters with respect to $(M,z)$ and found no strong mass dependence. However, we observed a monotonically decreasing redshift dependence of the amplitude parameter $s$, as illustrated in Figure~\ref{fig:cov_params_vary_mass_redshift}. To explain both the halo mass and the redshift dependence, we used the peak height of the halo, $\nu(M,z)$.

\begin{figure}
    \centering
    \includegraphics[width=\linewidth]{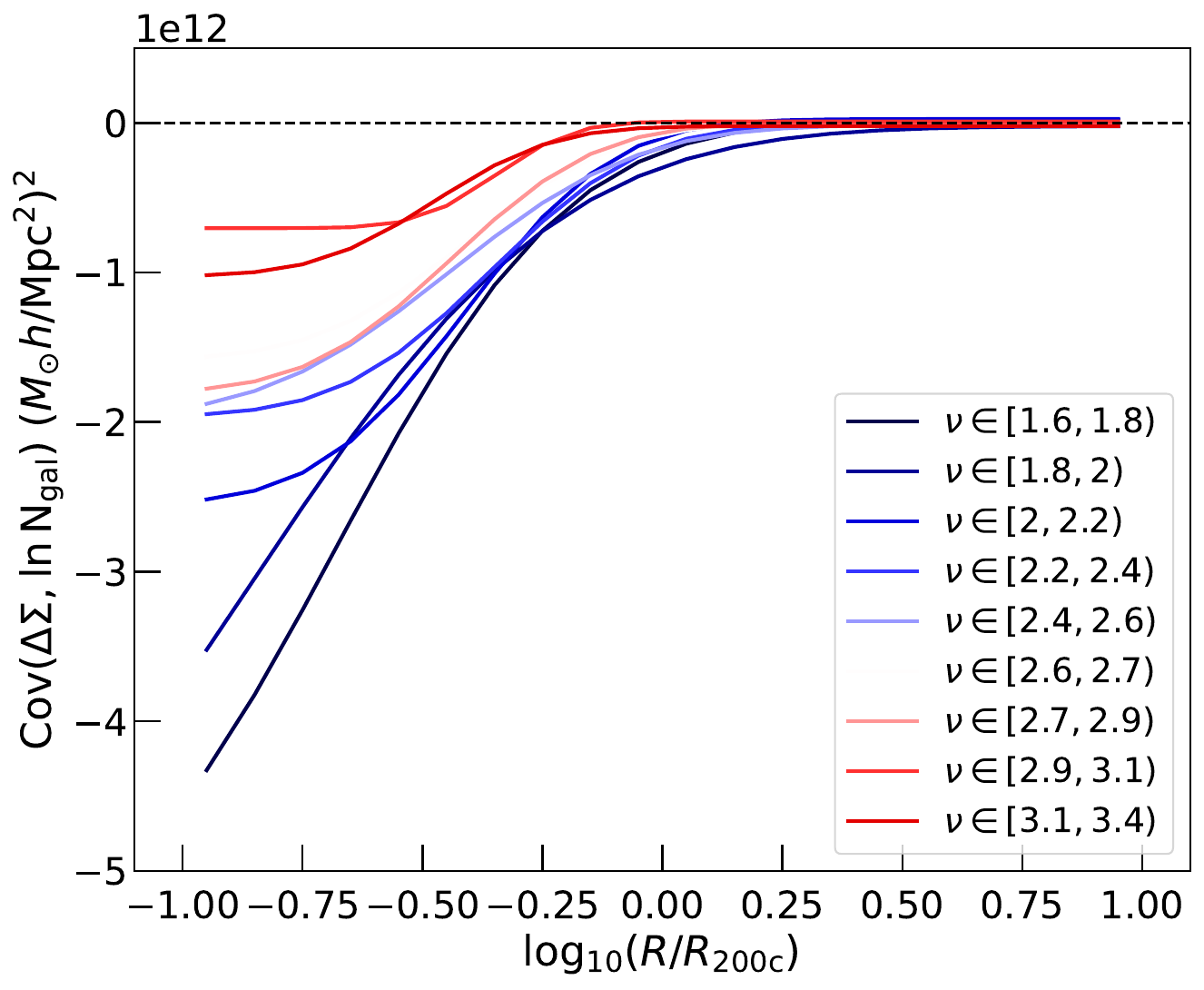}
    \caption{Best-fit ``full'' error function model for ${\rm Cov}(\Delta\Sigma, \ln\Ngal \mid M, z)$ when binned in deciles of halo peak height. The first nine bins reject the null hypothesis at a $p < 0.01$ level, and the highest decile rejects the null hypothesis at a $p < 0.05$ level. We can provide posterior constraints for all bins in peak height except for the one with the highest peak height value. } 
    \label{fig:cov_peakheight_measurement}
\end{figure}

\subsection{Binned by peak height}
\label{subsec:cov_binned_peakheight}

An alternative binning scheme that encapsulates both the halo mass and redshift information is to bin halos by the peak height parameter, defined as
\beqa
\nu = \frac{\delta_c}{\sigma(R,a)},
\eeqa
where $\delta_c(z)$ is the collapse overdensity at which gravitational collapses enter the non-linear regime and $\sigma(R,a)$ is the smoothing scale seen in Equation~\eqref{eqn:cosmic_variance} at the radius of the cluster. For an Einstein-de Sitter universe ($\Omega_m = 1$, $\Omega_\Lambda$ = 0) $\delta_c \approx 1.686$ at the epoch of collapse and is weakly dependent on cosmology and redshift \citep{Percival2005}. $\sigma(R,a)$ scales as $\sigma(M,a) = \sigma(M, a=1)D_{+0}(a)$ at the linear collapse regime, where $D_{+0}(a) \equiv D_{+}(a)/D_{+}(a=1)$. Here $D_{+}(a)$ is the linear growth factor defined as
\beqa
D_{+}(a) = \frac{5\Omega_M}{2}E(a)\int^a_0 \frac{da'}{[a'E(a')]^3},
\eeqa
for a $\Lambda$CDM cosmology, where $E(a) \equiv H(a)/H_0$ is the normalised Hubble parameter. $\sigma(R,z)$ depends strongly on redshift, and hence, the peak height $\nu$ strongly depends on the halo radius and the redshift of non-linear collapse. 

\begin{figure}
    \centering
    \includegraphics[width=\linewidth]{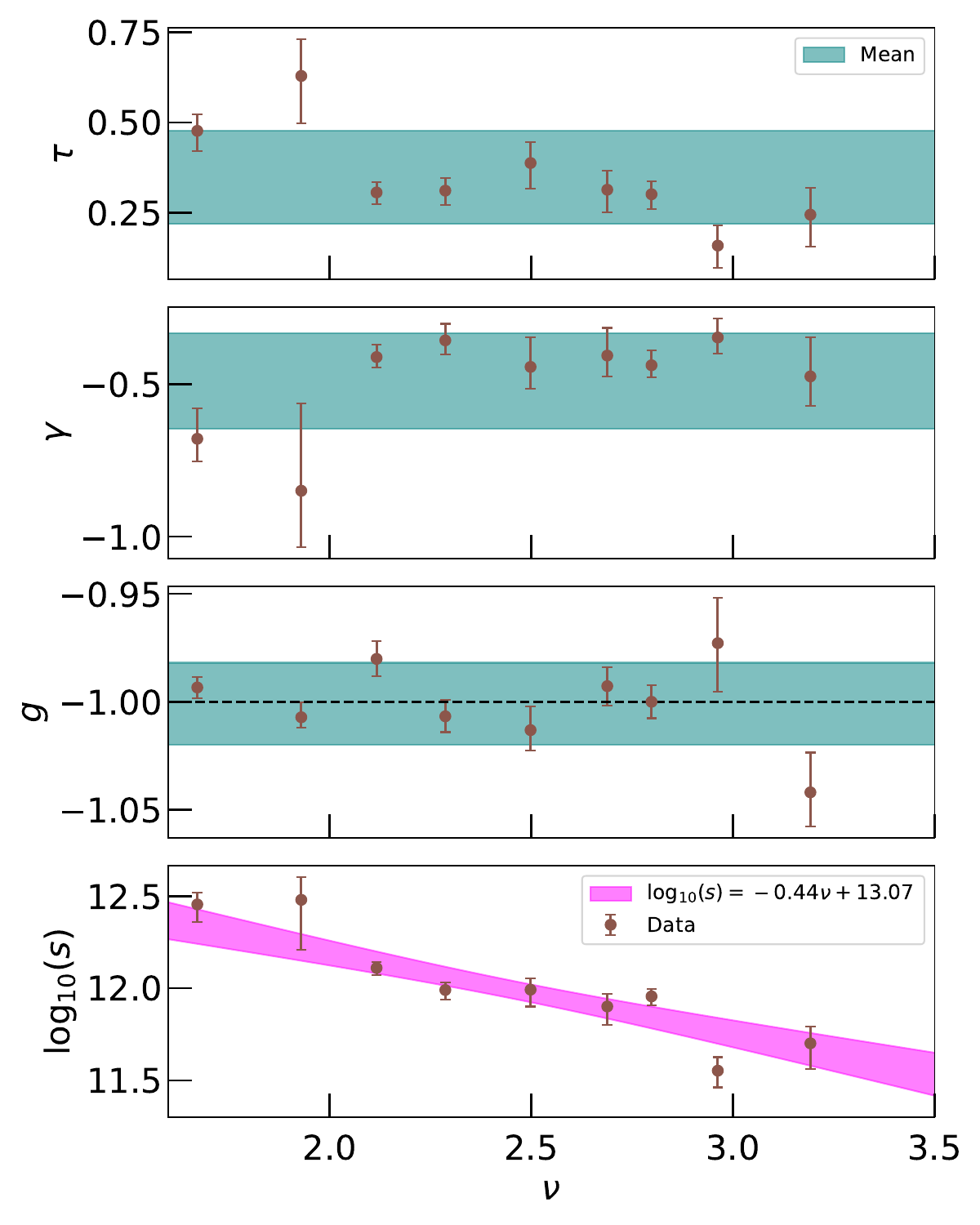}
    \caption{The evolution of shape parameters for ${\rm Cov}(\Delta\Sigma, \ln\Ngal \mid M, z)$ binned in deciles of the peak height $\nu$, excluding the highest decile. The parameters $\tau$, $\gamma$, and $g$ show little dependency with $\nu$ while the amplitude $s$ exhibits a log-linear relationship with $\nu$ of the form shown in Equation~\eqref{eqn:logs_nu}. The mean $g$ is consistent with -1. The horizontal teal bands fill the $1\sigma$ range around the mean, and the pink line is the best log-linear fit between $s$ and $\nu$ with $1\sigma$ confidence bands. } 
    \label{fig:param_peakheight}
\end{figure}

The peak height has been adopted to simplify the mass and redshift dependence in various halo properties, such as halo concentration \citep{Prada_2012} and halo triaxiality \citep{Allgood2006}. Here, we explore whether the peak height can serve as a universal parameter to explain the scale and shape of ${\rm Cov}(\Delta\Sigma, \ln\Ngal \mid M, z)$. We bin the halos into deciles of $\nu$ and set posterior constraints on the shape of the covariance using our \textit{erf} model in the full model case. In the highest decile (90\%-100\% percentile), we reject the null-correlation hypothesis at the $p =0.01$ level, but due to the size of the error bars, the shape of the parameters $\tau$, $\gamma$, and $g$ and largely unconstrained and $s = 10^{12} \times 0.16^{+0.34}_{-0.12}$. Due to the large degeneracy, we exclude the highest decile from our dataset and limit the range of our model to $\nu \in [1.57,3.40)$, which spans 0\% to 90\% of our sample set. The large error bars may be due to the fact that the halo abundance as a function of $\nu$ falls precipitously around $\nu \sim 4$, so the highest decile spans a wide tail of high $\nu \in [3.4, 4.6)$. The plots in Figure \ref{fig:cov_peakheight_measurement} are the best-fit templates when binned by peak height, and Figure~\ref{fig:param_peakheight} shows the best-fit parameters as a function of peak height. We do not see a strong dependence on the peak height for $\tau$, $\gamma$ and $g$. For $s$, its dependence on $\nu$ can be modeled as a log-linear relation of the form
\beqa
\log_{10}(s) = C_s + \alpha \nu,
\label{eqn:logs_nu}
\eeqa
with $C_s = 13.07^{+0.26}_{-0.26}$ and $\alpha = -0.44^{+0.11}_{-0.11}$. At the highest decile, the $s = 10^{12} \times 0.16^{+0.34}_{-0.12}$ falls within the $1\sigma$ confidence band of the log-linear fit. Compared to the first nine deciles, the fit yields a $\chi^2$ $p$-value of 0.73. The negative slope between $s$ and $\nu$ indicates that more massive halos at the cosmic era of their formation exhibit a lesser anti-correlation between $\Delta\Sigma$ and $\ln\Ngal$. 

\section{Impact of ${\rm Cov}(\Delta\Sigma, \ln\Ngal \mid M, z)$ on weak lensing measurements} \label{sec:mass_calibration}
\begin{figure*}
    \centering
    \includegraphics[width=\linewidth]{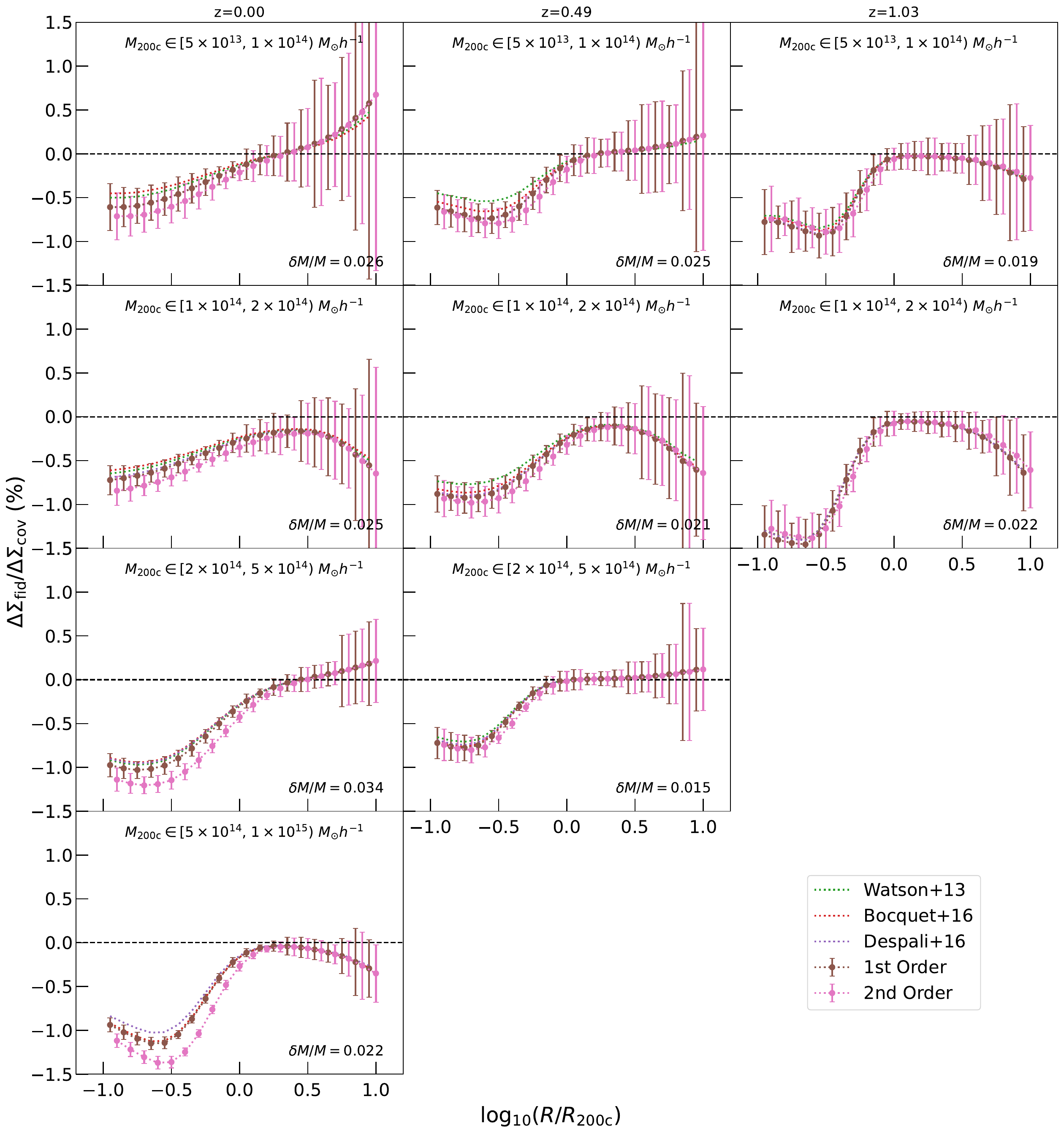}
    \caption{The percent level change in stacked $\Delta\Sigma$ measurements after including the covariance terms in Equations~\eqref{eq:1st_order_correction} and \eqref{eq:2nd_order_correction_s_b} as denoted by $\DS_{\rm cov}$ and without applying corrections as denoted by $\DS_{\rm fid}$. The slope and curvature of the halo mass function are calculated numerically from the \citet{Tinker_2008} halo mass function in our nominal correction. The errors are taken from bootstrapped errors of the covariance. We compare the results with first-order corrections from other halo mass functions using \citet{Watson_2013, Bocquet_2016, Despali_2016}. We find that the percentile difference in $\Delta\Sigma$ far exceeds the uncertainty in the choice of halo mass functions, and that second-order corrections are subdominant to the first-order correction itself, which is at a $\sim 1\%$ level at small scales for $\Delta\Sigma$ and propagates into an \textit{upward correction} of stacked halo mass of $\delta M / M \sim 2-3\%$ for most bins \textit{after} applying the correction.}
    \label{fig:DS_cov_correction}
\end{figure*}

To assess the impact of ${\rm Cov}(\Delta\Sigma, \ln\Ngal \mid M, z)$ on the scaling relation $\avg{\Delta\Sigma \mid N_{\rm gal}, z}$, we utilise Equation~\eqref{eq:1st_order_correction} for the first-order correction and Equation~\eqref{eq:2nd_order_correction_s_b} for the 2nd order. The mean mass of the halos in each $(M,z)$ bin is chosen as the pivot mass around which the HMF is Taylor expanded, and the intercept $\pi_{\Ngal}$ and slope $\alpha_{\Ngal}$ for the richness-mass scaling relation shown in Equation~\eqref{eq:model_richness} are computed locally at the pivot point in each bin $(M,z)$ bin. Our mock data, binned in $R_{\rm 200c}$, yields results that are consistent with the global richness-mass relation found in the literature \citep{Bocquet_2016, Costanzi_2021, To_2021}, as shown in Figure~\ref{fig:MOR_DMPL2_vs_data}.

For the correction terms, we adopt the \citet{Tinker_2008} halo mass function as our nominal model and compute the numeric log-derivative for values of $\gamma_1$ and $\gamma_2$, the log-slope and curvature of the halo mass function around the pivot mass. We compare the Tinker mass function results with others, including \citet{Watson_2013, Bocquet_2016, Despali_2016}, and find that the difference is subdominant to the first-order correction, which is at a $\sim 1\%$ level at small scales, as shown in Figure~\ref{fig:DS_cov_correction}.

To estimate the mass bias in each bin, we stack $\avg{\Delta\Sigma \mid M, z}$ and model the profiles as if they were individual halos with a mean mass, redshift and concentration as described in Equations \eqref{eqn:halo_matter_corr}-\eqref{eqn:xi_max}. We assume NFW profile using the concentration-mass model of \citet{Diemer_19} in the one-halo regime. The two-halo regime should not be affected, as the covariance is consistent with zero at $ R \gtrsim R_{\rm 200c}$. We convert the 3D overdensity of the modeled halo $\xi_{hm}$ to $\Delta\Sigma$ using Equations~\eqref{eqn:halo_matter_corr} and~\eqref{eqn:DeltaSigma}, and then apply the first-order correction in Equation~\eqref{eq:1st_order_correction}. Using a Monte Carlo method we obtain the expected mass with and without this correction and report the change in the mean halo mass with this correction for each $(M,z)$ bin. As shown in Figure~\ref{fig:DS_cov_correction}, we find that adding the correction leads to an \textit{upward correction} of the stacked halo mass of approximately $\delta M/M \sim 2-3\%$ for most $(M,z)$ bins.

\section{Explaining the covariance} 
\label{sec:Xparams_cov}

\subsection{Secondary Halo Parameter Dependence of $\ln\Ngal$}
\label{subsec:richness_Xparam}

We employed a multi-variable linear regression model to determine the best-fit when incorporating secondary properties in the regression. Initially, considering the full set of parameters listed in Table~\ref{tab:notation_Xparam}, we applied a backward modeling scheme to identify the relevant parameters of interest. Details of this process can be found in Appendix~\ref{sec:multilinear_modeling}, which led to the selection of the following secondary halo parameters for our model: $\Pi \subset \{\Gamma_{\rm 2dyn}, a_{\rm 1/2}, c_{\rm vir}, T/|U|, X_{\rm off} \}$. The resulting model demonstrated good explanatory power, as indicated by a high $R^2$ coefficient. Additionally, the model passed various tests, including variance inflation, global F-statistic, partial F-statistic, T-statistic, scatter heteroscedasticity, and scatter normality in most cases. Specifically, through a comparison of $F_{\rm partial}$ values, we found that richness could be modeled by a multi-linear equation involving all secondary halo parameters. Further information can be found in Table~\ref{tab:richness_multilinear_coeff}, where the F-statistic demonstrates that all parameters are statistically significant. Only when considered collectively can they accurately reflect the dependence of richness on halo formation history.

To establish informative priors for upcoming weak lensing surveys such as HSC and LSST, we examined whether the dependence of $N_{\rm gal}$ on secondary halo properties, as inferred from the slope $\beta_{\Ngal}$, aligns with arguments based on halo formation physics. We expected that $\beta_{{\Ngal, c_{\rm vir}}}$ resulting from the formation of satellite galaxies (equivalent to $N_{\rm gal}-1$ in the presence of a central galaxy) within halos would exhibit a negative relation, i.e., $\beta_{{\Ngal, c_{\rm vir} }} < 0$. Simulation-based studies have suggested that early-forming halos possess higher concentrations \citep{Wechsler_2002}, and correspondingly, high-concentration halos (which form early) have fewer satellite galaxies due to galaxy mergers within the halos \citep{Zentner_2005}. This effect is known as galaxy assembly bias \citep{Wechsler_2018} --- the change in galaxy properties inside a halo at fixed mass due to the halo formation history. There is marginal evidence of the existence of assembly bias from recent observations using galaxy clustering techniques \citep{Zentner_2019, Wang_2022}, as well as measurements of the magnitude gap between the brightest central galaxy (BCG) and a neighboring galaxy as a proxy for formation time \citep{Hearin_2013, Golden_Marx_2018, Farahi_2020}.

As noted in Table~\ref{tab:richness_multilinear_coeff}, the signs of $\beta_{\Ngal,i}$ for the remaining parameters $i \in \{a_{1/2}, T/|U|, \Gamma_{\rm 2dyn}, X_{\rm off} \}$ align with our expectations of assembly bias in most bins --- late-forming clusters undergo more rapid mass accretion (higher $\Gamma_{\rm 2dyn}$) and are less virialized (higher $\rm{T/|U|}$), and because they also from the galaxy assembly bias mentioned above are richer in galaxy number counts when conditioned on the mass, we expect a positive partial slope $\beta_{\Ngal,\Gamma_{\rm 2dyn}}$ and $\beta_{\Ngal, \rm{T/|U|}}$. The case for $a_{1/2}$ and $X_{\rm off}$ is more complicated. Under the isolated formation of halos $a_{1/2}$ and $X_{\rm off}$ would be smaller for earlier forming halos due to the monotonic mass accretion and relaxation of halos over long time scales. However, as halos undergo mergers and tidal stripping the monotonicity of the parameters over time is not guaranteed. Therefore, we see a mixture of positive and negative partial slopes $\beta_{\Ngal,\rm{a_{1/2}}}$ and $\beta_{\Ngal,\rm{X_{\rm off}}}$ in these cases. To describe the physical mechanisms on a case-by-case basis would require that we probe into the halo merger tree history of individual halos.

In this paper, we take a closer look at the sign of $\beta_{{\Ngal, c_{\rm vir} }}$ and observe that while the partial scope matches our expectations in most bins, in some mass bins at medium and high redshifts it changes signs from negative at lower redshifts to positive at higher ones. While we observe a diminishing impact of secondary halo properties on richness (indicated by a smaller absolute value for $\beta_{\Ngal, c_{\rm vir}}$), the reversal of the coefficient's sign cannot be solely attributed to statistical fluctuations around zero, as some values are inconsistent with zero at levels exceeding $3\sigma$.

This issue can be attributed to the effect of major mergers on concentration. Recent studies \citep{Ludlow_2012, Wang_2022, Lee_2023} have shown that halos, during major merger events, experience a transient fluctuation in concentration before returning to the mean relation over a time period slightly less than the dynamical time of the halo. The measured concentration spike during major mergers, particularly prominent at higher redshifts, could explain a positive $\beta_{\{\Ngal, c_{\rm vir} \}}$.

To test this hypothesis, we employ a toy model that divides halos in each $(M,z)$ bin based on the median $\Gamma_{\rm 1dyn}$ into low-$\Gamma_{\rm 1dyn}$ and high-$\Gamma_{\rm 1dyn}$ subsamples. Given the timescale of mergers to be roughly the dynamical time of the halo, we choose $\Gamma_{\rm 1dyn}$ as a good proxy for potential merger events even though this parameter is excluded in the final linear regression model due to multicollinearity (see Appendix \ref{sec:multilinear_modeling}.)

Figure~\ref{fig:richness_concentration_MAH} displays the halo concentration plotted against the richness residuals, separated by $\Gamma_{\rm 1dyn}$, at benchmark bins of $M_{\rm 200c} \in [5 \times10^{13}, 1\times10^{14})\ M_{\odot}h^{-1}$ at three different redshift snapshots of $z=0, 0.49, 1.03$. At $z=0$, we observe a negative slope as expected from halo formation physics for both low-$\Gamma_{\rm 1dyn}$ and high-$\Gamma_{\rm 1dyn}$ subsamples, as well as for the overall sample. Furthermore, we observe a change in the slope between the low-$\Gamma_{\rm 1dyn}$ and high-$\Gamma_{\rm 1dyn}$ sub-samples, which can be explained by the gradual increase (or decrease) in concentration ($\Gamma_{\rm 1dyn}$) over time, even without major merger events \citep{Wechsler_2002, Zhao_2003, Lu_2006}. At redshifts of $z=0.49, 1.03$, we observe a positive slope in the overall and/or high-$\Gamma_{\rm 1dyn}$ samples, which contradicts the scaling relations between HOD and concentration in models that track their gradual evolution over $T \gg T_{\rm 1dyn}$. However, in the presence of major mergers, when $\Gamma_{\rm 1dyn}$ is significantly enhanced, the halo concentration may also experience a transient spike after the merger. The deviation from hydrostatic equilibrium provides a \textit{plausible} explanation for a positive $\beta{\{\Ngal, c_{\rm vir} \}}$, which could be fully tested on MDPL2 through the reconstruction of halo merger trees, an analysis beyond the scope of this paper.

\begin{figure*}
    \centering
    \includegraphics[width=1\linewidth]{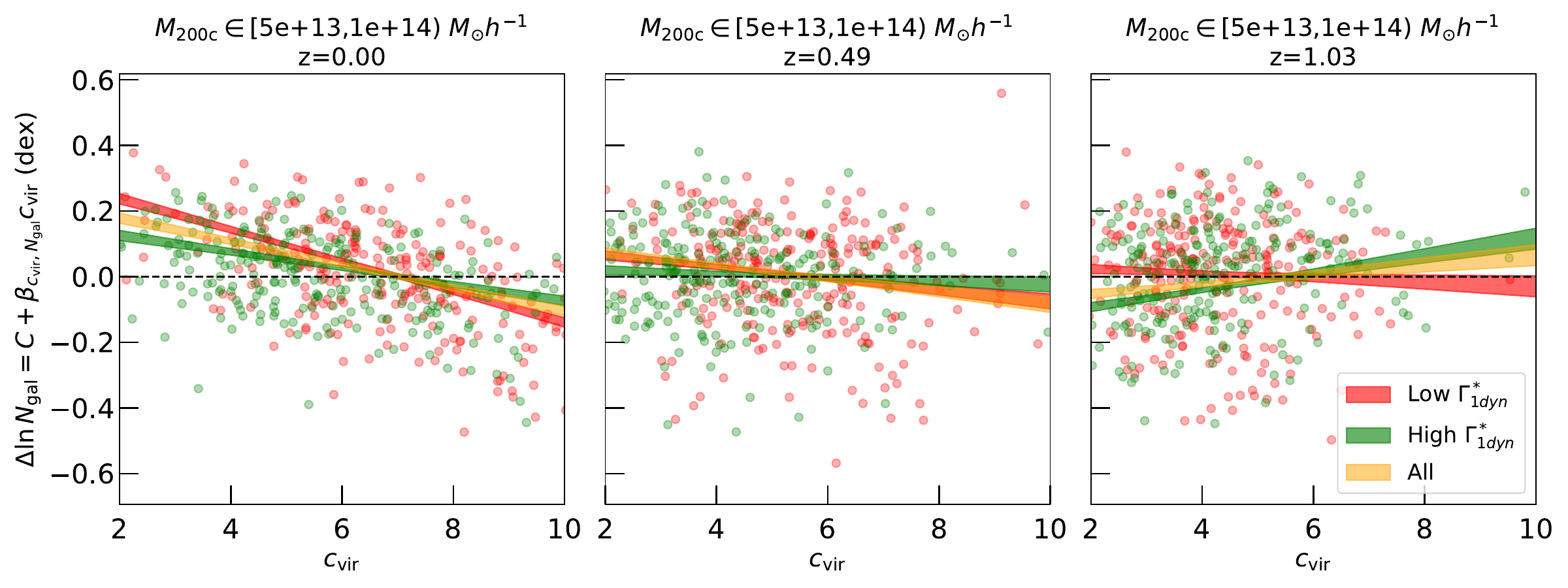}
    \caption{Residual log-richness versus concentration relation in subsets of halo mass accretion rate (MAR). The figure consists of three panels --- left, middle, and right panels corresponding to $z=0$, $z=0.49$, and $z=1.03$, respectively, all on a benchmark mass bin of $M_{\rm 200c} \in [5\times10^{13}, 1\times10^{14}) M_{\odot}h^{-1}$. The sample is split into low and high $\Gamma_{\rm 1dyn}$ based on their median values. The scatter plot illustrates the data points, while the shaded regions show the best-fit linear fit with $1\sigma$ confidence interval for the main sample and each sub-sample. At $z=0$, the richness-concentration relation exhibits a negative slope, consistent with our expectations of halo formation physics. The slopes for the low and high $\Gamma_{\rm 1dyn}$ subsamples diverge due to the negative correlation between concentration and MAR. However, at $z=0.49$ and $z=1.03$, the slopes for the entire sample and/or the high $\Gamma_{\rm 1dyn}$ subsample become positive, contrary to our observations of the richness-concentration relation. In contrast, the low $\Gamma_{\rm 1dyn}$ subsample still shows a negative slope. These findings suggest that at medium to high redshifts, a subset of unrelaxed and recently merged halos with high MAR could elevate the concentration from its expected value at hydrostatic equilibrium.}
    \label{fig:richness_concentration_MAH}
\end{figure*}

\subsection{Secondary Halo Parameter Dependence of $\Delta\Sigma$}
\label{subsec: DS_Xparam}

\begin{figure*}
    \centering
    \includegraphics[width=\textwidth]{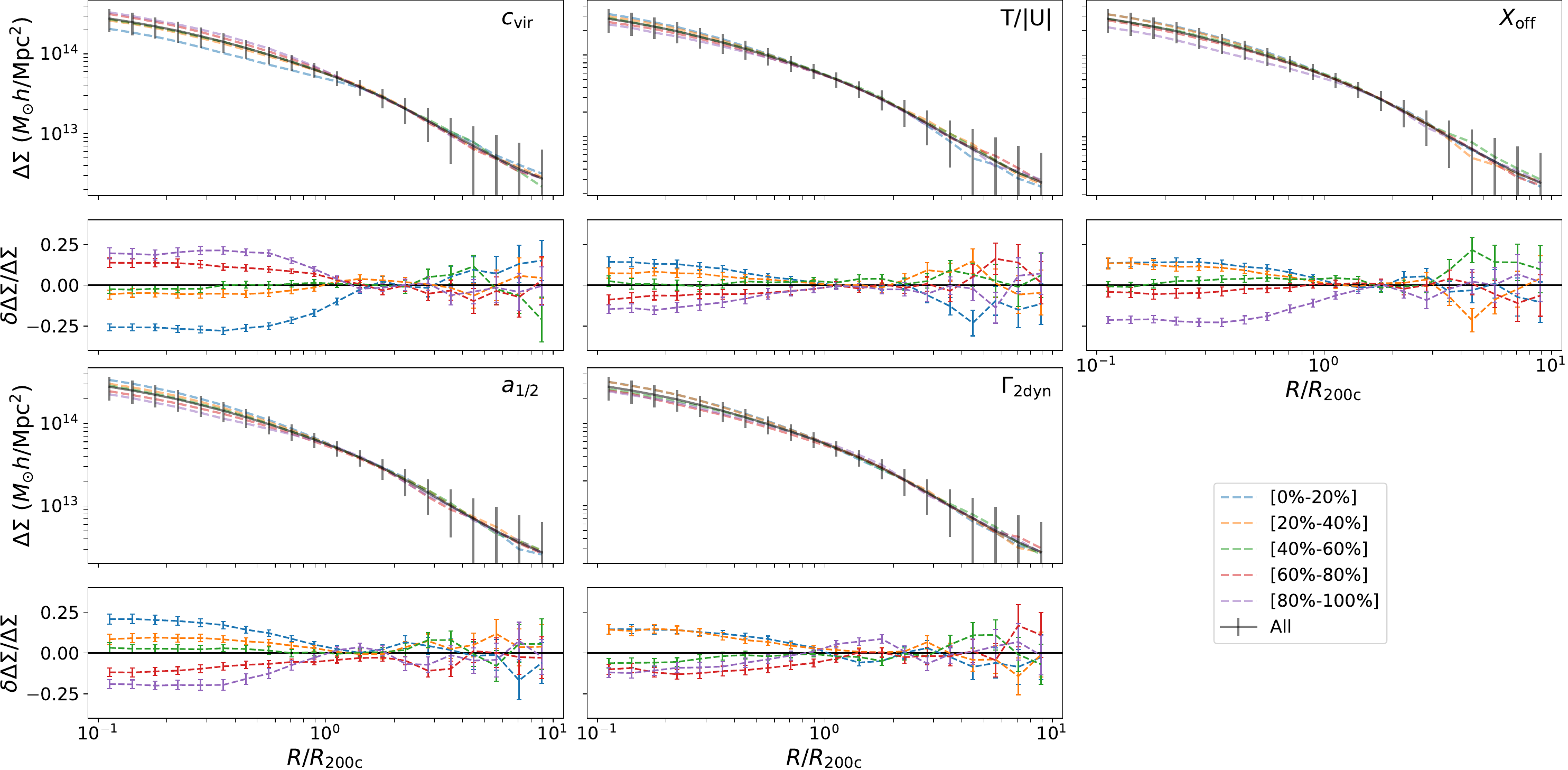}
    \caption{The dependence of $\Delta\Sigma$ on accretion history parameters in $M_{\rm 200c} \in [2\times10^{14}, 5\times10^{14}) M_{\odot}h^{-2}$,~$z=0$. In each of the 5 panels is plotted the $\Delta\Sigma$ in different quintiles of the accretion history parameter $\Pi \in \{a_{1/2}, c_{\rm vir}, T/|U|, \Gamma_{\rm 2dyn}, X_{\rm off}\}$ compared to the mean $\Delta\Sigma$. For $\{a_{1/2}, T/|U|, \Gamma_{\rm 2dyn}, X_{\rm off}\}$ there is a strong negative correlation at small scales at $R \lesssim R_{\rm 200c}$ and for $c_{\rm vir}$ we find a strong positive correlation at $R \lesssim R_{\rm 200c}$. Comparing the width of $\Delta\Sigma$ binned at different quintiles of $\Pi$ with the standard deviation of the profiles, we find that accretion history parameters account for much of the variance at small scales and play a negligible role at large scales. }
    \label{fig:DS_accretion_history}
\end{figure*}

In this section, we employ a multi-linear regression approach to model the lensing signal, similar to the methodology described in \S\ref{subsec:richness_Xparam}. We extend this approach to the model $P(\Delta\Sigma| \Ngal(\Pi), M, z)$ as a linear function of $\Pi$. Upon analysing different $(M, z, r_p)$ bins, we observe that the reduced parameters $\Pi \subset \{a_{1/2}, c_{\rm vir}, T/U, \Gamma_{\rm 2dyn}, X_{\rm off} \}$ pass the variance inflation factor (VIF) test for multi-collinearity, or in other words we showed that the variance is not inflated and thereby made less reliable in the case that the secondary halo parameters in the full model are highly correlated. As with the case for the lensing signal, this indicates that $\ln\Ngal$ can be described without redundancy by a linear decomposition of these reduced parameters. Furthermore, most bins exhibit homoscedasticity, as confirmed by passing the Breusch-Pagan Lagrange multiplier test. This implies that the scatter terms $\sigma_{\DSig}$ and $\sigma_{\Pi_i}$ remain constant within each bin, with a few exceptions. Lastly, the scatter $\sigma_{\DSig|\Ngal}$ in most bins (with a few exceptions) meets the criteria of the Shapiro-Wilk test for Gaussianity, suggesting that the distribution closely resembles a Gaussian distribution.

The multi-linear regression is a good fit to the conditioned lensing signal if we assume that $P(\ln\Ngal | M, z)$ and $P(\Delta\Sigma | M, z)$ can be modeled with a normal distribution \citep[e.g.,][]{Anbajagane:2020stellar, Costanzi_2021, To_2021}. In this case $P(\Delta\Sigma \mid \Ngal,  M, z)$ is a multi-linear equation with respect to the secondary halo parameters with mean
\begin{align}
\label{eq:DS_given_lambda_Xparam_mean}
\avg{\Delta\Sigma | \Ngal, M, z } ~=&~ \avg{\Delta\Sigma|N_{\rm gal}, M, z} \\ \nonumber 
+& C_1\sigma_{\Delta\Sigma}\Big(\sum_i \frac{\beta_{\Ngal,i} }{\sigma_{\Pi_i}}\rho_{\Delta\Sigma-\Pi_i} \times (\Pi_i - \avg{\Pi_i | M, z})\Big)
\end{align}
and is normally distributed around the mean with variance
\begin{align}
\label{eq:DS_given_lambda_Xparam_var}
\qquad \sigma^2_{\Delta\Sigma|\ln{\Ngal}} \quad =&~  \sigma^2_{0} +  C_2\sum_i{\beta_{\Ngal,i}^2 \sigma^2_{\Pi_i}(1-\rho^2_{\Delta\Sigma-\Pi_i})}+ \\ \nonumber
& C_3\sum^{j\neq i}_{i,j}{\rho_{\Pi_i-\Pi_j}\sigma_{\Pi_i}\sigma_{\Pi_j}}.
\end{align}
The parameters $C1$, $C2$, $C3$, and $\sigma_0$ can be explicitly derived where $P(\ln\Ngal | M, z)$ and $P(\Delta\Sigma | M, z)$ are known, but the exact values are not essential for this paper. We refer the reader to Appendix \ref{sec:proof_DS_given_lambda} for derivations of Equations \eqref{eq:DS_given_lambda_Xparam_mean} \& \eqref{eq:DS_given_lambda_Xparam_var}. 

We note that only in bins of $R \lesssim R_{\rm 200c}$ do the multilinear regression models pass the global F-statistic test and the T-statistic test for each parameter. This result suggests that, at $R \gtrsim R_{\rm 200c}$, we find little correlation between $\Delta\Sigma$ and $\Pi_i$. Because the scatter still passes the Shapiro-Walk test for Gaussanity, the conditional probability $P(\Delta\Sigma | N_{\rm gal}, M, z)$ at large scales is still normally distributed, but with $\rho_{\DSig-\Pi_i} = 0$. 
By setting $\rho_{\DSig-\Pi_i} = 0$ the variance can be reduced to
\begin{equation}
\sigma^2_{\DSig|\Ngal} \quad =  \sigma^2_{0} +  \sum_i{\beta_{\Ngal,i}^2 \sigma^2_{\Pi_i}} +\sum^{j\neq i}_{i,j}{\rho_{\Pi_i-\Pi_j}\sigma_{\Pi_i}\sigma_{\Pi_j}}.
\end{equation}

We visualise the dependence of $\rho_{\Delta\Sigma-\Pi_i}$ on $R/R_{\rm 200c}$ in Figure~\ref{fig:DS_accretion_history}. By dividing $\Delta\Sigma$ into quintiles of $\Pi_i$ we find a strong correlation for all parameters at $R \lesssim R_{\rm 200c}$ and a null correlation at $R \gtrsim R_{\rm 200c}$. On small scales, our results show a positive correlation for concentration and a negative correlation for $\{a_{1/2}, T/U, \Gamma_{\rm 2dyn}, X_{\rm off} \}$. We observe that this trend holds for all $(M,z)$ bins plotted for a benchmark bin of $M_{\rm 200c} \in [2\times10^{14}, 5\times10^{14}) M_{\odot}h^{-1}$ at $z=0$.

The dependence of $\Delta\Sigma$ on secondary halo parameters qualitatively agrees with \citet{Xhakaj2022} wherein they targeted a narrow mass bin, with residual mass dependency inside the bin resampled so that mass follows the same distribution. In our work, we remove the mass dependency with the KLLR method \citep[][]{Farahi:2022KLLR}, which achieves the same effect. We extend their results to mass and redshift bins probed by the optical surveys and quantitatively show that the dependence of $\Delta\Sigma$ on $\Pi$ can be modeled as a multi-linear equation.

\subsection{Results: Secondary Halo Parameter Dependence of ${\rm Cov}(\Delta\Sigma, \ln{\Ngal} \mid M, z)$}

\begin{figure*}
    \centering
    \includegraphics[width=\linewidth]{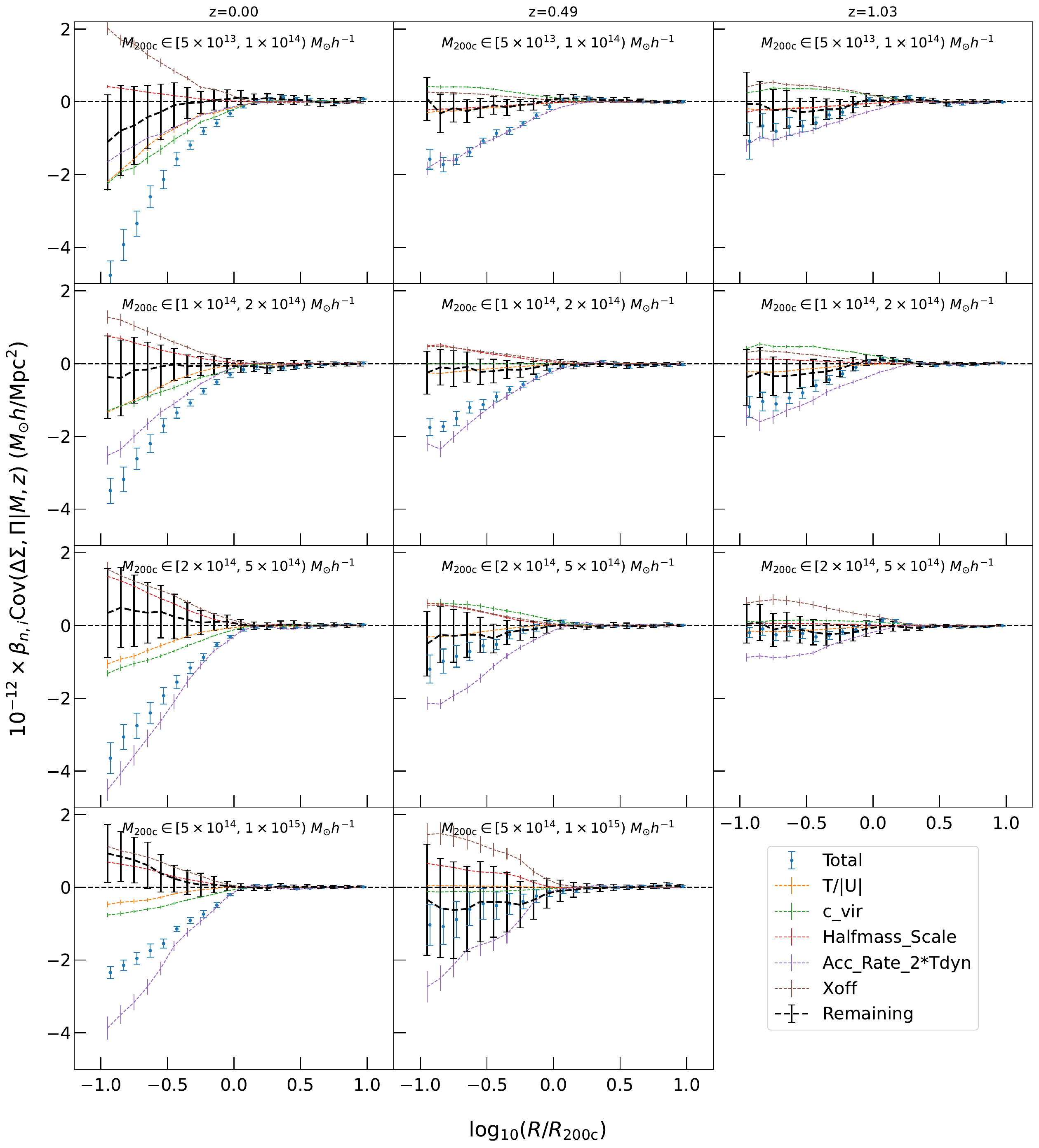}
    \caption{The dependence of ${\rm Cov}(\DSig,\ln\Ngal)$ on secondary halo parameters $\Pi$. The solid blue line is the total covariance. The dashed lines of the lines represent the covariance contribution coming from each of the secondary halo parameters modeled and governed by Equation~\eqref{eqn:cov_hypothesis}, where ${\rm Cov}(\Delta\Sigma, \Pi | M, z)$ comes from the dependence of $\Delta\Sigma$ and the slope $\beta_i$ is the dependency of richness. The thick black dashed line is the remaining covariance after removing the contribution from each $\Pi_i$ term; the errors are obtained by adding the total and individual errors in quadrature without considering the correlations between terms. In agreement with our hypothesis in Equation~\eqref{eqn:cov_hypothesis}, the remaining term is consistent with null at a $p<0.01$ level for all bins. }
    \label{fig:cov_Xparams_split}
\end{figure*}

In Figure~\ref{fig:cov_Xparams_split}, we observe that the total covariance ${\rm Cov}(\DSig, \ln \Ngal| M, z)$, which remains after removing the contribution of each secondary halo parameter $\beta_{\Ngal,i}{\rm Cov}(\DSig, \Pi_i| M, z)$, is consistent with zero at a significance threshold of $0.05$ in all bins. The errors on the total covariance and individual contributions are computed using bootstrapping, and the errors on the remaining term are determined by adding the errors of the total and individual terms in quadrature.

Based on our hypothesis in Equation~\eqref{eqn:cov_hypothesis}, we conclude that the set of secondary halo parameters $\Pi$, which are related to the formation time and the mass accretion history of the halos, can fully explain the joint distribution of $\DSig$ and $\ln \Ngal$ given the precision allowed by current errors, limited by the resolution limit (see Appendix \ref{sec:ptcl_resolution} for information on particle resolution and measurement errors).

Since the joint distribution of $\DSig$ and $\ln \Ngal$ follows a multivariate normal distribution, $P(\DSig, \ln \Ngal | M, z)$ is completely characterised by its mean relation and ${\rm Cov}(\DSig, \ln\Ngal \mid M, z)$. It should be noted that the contribution of each individual parameter to the total covariance, $\beta_{\Ngal, i}{\rm Cov}(\DSig, \Pi_i \mid M, z)$, is determined by the richness dependency captured by the slope $\beta_{\Ngal, i}$ and the $\Delta\Sigma$ dependency represented by ${\rm Cov}(\DSig, \Pi_i \mid M, z)$. Qualitatively, individual contributions to total covariance maintain their sign when both $\Delta\Sigma$ and $\Ngal$ contributions preserve their sign. Consistent with the arguments of halo formation, $\Pi$ correlates with $\Delta\Sigma$ at small scales, as demonstrated in Figure~\ref{fig:DS_accretion_history} across all $(M,z)$ bins. In most cases, the dependence of the richness on secondary halo parameters also maintains its sign across the $(M,z)$ bins. In instances where we encounter a sign reversal in the $\ln\Ngal-c_{\rm vir}$ relation, we speculate that it is due to a transient increase in concentration following a major merger.

Furthermore, the total and individual contributions to the covariance tend to decrease in magnitude at smaller scales with increasing redshift. This decrease in covariance can be attributed to two factors: the decreasing explanatory power of $\Pi$ on richness, as indicated by the decreasing values of $R^2$ and $F_{\rm partial}$ values in Table \ref{tab:richness_multilinear_coeff}, and the decreasing absolute value of ${\rm Cov}(\DSig, \Pi_i \mid M, z)$. This trend aligns with the idea that as halos have more time to form, the secondary halo properties related to the mass accretion history become more significant both in richness and in $\Delta\Sigma$. As discussed in \S\ref{sec:cov_form}, the dependence of the mass and redshift on covariance can be explained by the halo peak height, $\nu(M,z)$.

\section{Discussions} \label{sec:discussion}

\paragraph*{Intrinsic vs. Extrinsic Covariance.}
Our study distinguishes between the intrinsic covariance investigated here and that observed in empirical cluster data sets. This distinction arises from systematic biases introduced by the cluster-finding algorithm (extrinsic component) and the underlying physics governing halo formation (intrinsic component). Specifically, our analysis involves counting galaxies in 3D physical space. In contrast, a realistic cluster finder like redMaPPer \citep{Rykoff:2014RedMapper} employs a probabilistic assignment of galaxies to halos in 2D physical space, considering projected radii and redshift through color matching onto the red sequence. Our study does not account for the observational systematics in redMaPPer associated with uncertainties in photometric redshifts and projection effects \citep{Rozo:2015RM-IV,Farahi_2016}.

We find that the fractional amplitude of the bias and the scale dependence on the lensing signal observed in our results (Fig. \ref{fig:DS_cov_correction}) are comparable to those reported by \citet{Farahi_2022_corr}, who measured the covariance between the dark matter density and the galaxy number count enclosed inside a halo after applying a realistic stellar mass cut. The enclosed mass within a 3D radius using the IllustrisTNG100 simulation is anchored at $z=0.24$. By comparing our findings to those obtained using a realistic cluster finder such as redMaPPer, we can unravel intrinsic and extrinsic contributions to the covariance between weak lensing observables \citep{Wu_2022}. This work provides more profound insights into the distinct effects originating from the underlying physics and the methodology employed in cluster-finding algorithms \citep{Euclid2019:cluster-comparison}.

\paragraph*{Projection Effects.}
A noteworthy distinction arises regarding the covariance observed in our study compared to others examining clusters in projected space. Projection effects can potentially introduce a sign flip in covariance, as they can positively bias both $\Delta\Sigma$ and richness \citep{Costanzi_2019, Wu_2022, Zhang22, Zhang_2023}. Particularly, \citet{Zhang_2023} showed that the lensing signal can be affected both at large and small scales from the preferential alignment of halo orientation with the underlying large-scale structure filament. \citet{Wu_2022} detected a positive correlation between $\Delta\Sigma$ and $\ln\Ngal$ employing Buzzard simulations, where $\Delta\Sigma$ were measured using dark matter particles and galaxy counts were performed within a cylindrical region of depth $60$~Mpc/h. Their investigation revealed that the positive correlation primarily stems from galaxy number counts beyond the halo's virial radius and within $60$~Mpc/h.
On the other hand, using the Dark Quest emulator and HOD-based galaxy catalogues \citep{Nishimichi19}, \citet{Sunayama_2020} found negligible deviations from the mean relation at small scales and an overall reduction in selection bias at large scales, approximately halving the effect observed by \citet{Wu_2022}. Furthermore, \citet[][]{Huang22}, using data from the Subaru Hyper Suprime-Cam (HSC) survey, observed that the selection bias is most prominent in the vicinity of the transition from the one-halo to the two-halo regime, as evidenced by the comparison between the outer stellar mass proxy and richness. In a study based on the IllustrisTNG300 simulation, \citet{Zhang22} discovered a net positive correlation between the fitted weak lensing mass and the projected 2D number count of the halo when conditioned on the halo mass.

These results suggest that projection effects can potentially introduce a positively correlated bias to both $\Delta\Sigma$ and $\Ngal$. We can estimate the impact of projection effects by comparing the intrinsic covariance measured in our study with the total covariance observed in the projected space. Consequently, our results serve two essential purposes: (i) elucidating the physical origins of the negative covariance and (ii) discerning intrinsic and extrinsic components to determine the covariance attributable to projection effects accurately.

\paragraph*{Radial Dependence.} There is a notable difference in the reported amplitude and scale dependence of covariance, which can be attributed to discrepancies in the employed halo occupation density models. Notably, a distinctive scale dependence discrepancy exists between simulation-based investigations of projection effects \citep{Sunayama_2020, Wu_2022, Salcedo19} and observational data from the HSC \citep{Huang22}. In particular, the analysis of observational data reveals a prominent 1 Mpc bump, which could be explained by uncertainties inherent in observations, such as miscentering effects. It is crucial to gain insight into the sensitivity of the covariance with respect to the model parameters and the influence of selection effects. Understanding these factors is essential to comprehensively interpret and account for the observed covariance in galaxy cluster survey studies.

\paragraph*{Accuracy vs. Precision.} The statistical power of current and future surveys enables us to determine the normalisation and slope of the mass--observable relations at a few percent levels \citep[e.g.,][]{Farahi_2016,Mantz2016WtG-V,Mulroy2019,To_2021}. However, these estimates are susceptible to known and unknown sources of systematic errors that inflate the uncertainties. These uncertainties introduce biases and degrade the accuracy of the results. Therefore, it is essential to carefully identify, quantify, and account for these systematic effects to ensure robust and reliable measurements. In this work, we focus on studying one of these sources of systematic uncertainty that was not considered previously.

\section{Summary} \label{sec:summary}

This work reveals insights into the scale-dependent covariance between weak lensing observables and the physical properties of the halo. Using the MDPL2 N-body simulation with galaxies painted using the SAGE semi-analytic model, we present several key findings:
\begin{itemize}
    \item We observe that the intrinsic covariance between $\Delta\Sigma$ and $\ln\Ngal$ enclosed within a 3D radius is negative at small scales and null at large scales in $(\ln M,z)$ ranges that cover optical surveys.
    
    \item We model the shape of the covariance across all bins using an error function that is insensitive to the radius definition used to define halo boundaries. 
    
    \item We find that the magnitude of the covariance is relatively insensitive to mass and decreases considerably with increasing redshift. The $(M,z)$ dependence of the shape of the covariance can be encapsulated by the peak height parameter $\nu(M,z)$, which suggests that the scale of the covariance is related to the formation history of halos.
    
    \item We show that incorporating the covariance into $\langle \Delta\Sigma| \Ngal, z, r_p \rangle$ using the first-order expansion of the halo mass function yields about $> 1\%$ bias on $\langle \Delta\Sigma | \Ngal, z, r_p \rangle$ at small scales, which implies a mass bias of $> 2 \%$ in the halo mass estimates in most bins.
    
    \item Our analysis reveals that the covariance between $\ln\Ngal$ and $\Delta\Sigma$ can be fully explained by secondary halo parameters related to the history of the halo assembly. This finding provides strong evidence that the non-zero covariance results from the variation in the formation history of dark matter halos.
    
\end{itemize}

This work underscores the importance of accounting for covariance in cluster mass calibration. Incorporating the covariance between richness and the weak lensing signal and its characterisation should be an essential component of weak lensing cluster mass calibration in upcoming optical cluster surveys. The results of this work can be introduced as a simulation-based prior for the forward modeling of scaling relations used by cluster cosmological analysis pipelines. Within the LSST-DESC framework, this systematic bias would be implemented in the CLMM pipeline \citep{Aguena_2021} to update the stacked weak lensing mass in each richness bin.  Considering this covariance paves the path toward percent-level accuracy cosmological constraints, thereby enhancing the precision and reliability of our scientific conclusions. Moving forward, it is imperative to integrate this understanding into the design and analysis of future optical cluster surveys.

\section*{Acknowledgements}
Author contributions are as follows: \\
Z. Zhang -- conceptualisation, code development, methodology, writing, edits \\
A. Farahi -- conceptualisation, code development, methodology, writing, edits, supervision \\
D. Nagai -- methodology, writing, edits, supervision \\
E. Lau -- methodology, code development, writing, edits, supervision \\
J. Frieman -- edits, supervision \\
M. Ricci -- edits, supervision, administrative organisation \\
A. Linden -- internal reviewer \\
H. Wu -- internal reviewer \\

We thank the anonymous referee for careful edits that greatly improved the quality of this manuscript. This paper has undergone an internal review in the LSST Dark Energy Science Collaboration. We thank Anja von der Linden, Tamas Varga and Hao-Yi Wu for serving as the LSST-DESC publication review committee. In addition, the authors thank Andrew Hearin, Heather Kelly, Johnny Esteves, Enia Xhakaj, and Conghao Zhou for their helpful discussions. 

The DESC acknowledges ongoing support from the Institut National de 
Physique Nucl\'eaire et de Physique des Particules in France; the 
Science \& Technology Facilities Council in the United Kingdom; and the Department of Energy, the National Science Foundation, and the LSST Corporation in the United States.  DESC uses resources of the IN2P3 Computing Center (CC-IN2P3--Lyon/Villeurbanne - France) funded by the Centre National de la Recherche Scientifique; the National Energy Research Scientific Computing Center, a DOE Office of Science User Facility supported by the Office of Science of the U.S.\ Department of Energy under Contract No.\ DE-AC02-05CH11231; STFC DiRAC HPC Facilities, funded by UK BEIS National E-infrastructure capital grants; and the UK particle physics grid, supported by the GridPP Collaboration.  This 
work was performed in part under DOE Contract DE-AC02-76SF00515.

The CosmoSim database used in this paper is a service by the Leibniz-Institute for Astrophysics Potsdam (AIP). The MultiDark database was developed with the Spanish MultiDark Consolider Project CSD2009-00064.

The authors gratefully acknowledge the Gauss Centre for Supercomputing e.V. (www.gauss-centre.eu) and the Partnership for Advanced Supercomputing in Europe (PRACE, www.prace-ri.eu) for funding the MultiDark simulation project by providing computing time on the GCS Supercomputer SuperMUC at Leibniz Supercomputing Centre (LRZ, www.lrz.de). The Bolshoi simulations have been performed within the Bolshoi project of the University of California High-Performance AstroComputing Center (UC-HiPACC) and were run at the NASA Ames Research Center.

This research used resources of the National Energy Research Scientific Computing Center, a DOE Office of Science User Facility supported by the Office of Science of the U.S. Department of Energy under Contract No. DE-AC02-05CH11231 using NERSC award HEP-ERCAP0022779.

DN acknowledges support by NSF (AST-2206055 \& 2307280) and NASA (80NSSC22K0821 \& TM3-24007X) grants. AF acknowledges the Texas Advanced Computing Center (TACC) at The University of Texas at Austin for providing HPC resources that have partially contributed to the research results reported within this paper. 

\section*{Data Availability}
Raw data from the MDPL2 simulation are publicly available at CosmoSim\footnote{https://www.cosmosim.org/}. Derived data products are available upon request. 
Our code is made publicly available at \url{https://github.com/BaryonPasters/BP_CorrelatedScatter_MDPL2}.


\bibliographystyle{mnras}
\bibliography{reference} 

\appendix
\section{Functional form}
\label{sec:functional_form}
This section aims to characterize the shape of the covariance across mass and redshift bins by fitting a template curve. The process involves several transformations and adjustments. First, a logarithmic transformation is applied to the radial bins, denoted as $x = \log_{10}(R/R_{\rm 200c})$. Then, a horizontal offset is introduced using a parameter $\gamma$, and scaling is applied using a parameter $\tau$. This results in a transformed variable $\tilde{x} = (x - \gamma)/\tau$.

To analyze the transformed data vector $f(\tilde{x})$, we test a set of functional forms presented in Table~\ref{tab:cov_functional_form}. The normalization factors and coefficients associated with these functions are chosen in such a way that $f(\tilde{x})$ approaches 1 at large scales, -1 at small scales, $f'(0)=1$ and $f(0) = 0$. 

A linear transformation of $f(\tilde{x})$ is then performed, given by $s\big(f(\tilde{x}) + g \big)$, where $g$ represents a vertical shift and $s$ represents a scaling factor. The magnitude of $s$ is comparable to the magnitude of $\text{Cov}(\Delta\Sigma, \ln\Ngal \mid M, z)$, while the parameters ${\gamma, \tau, g}$ are of the order of unity. These parameters, along with $s$, form the set of parameters denoted as $\boldsymbol{\theta} \in \{\tau,\gamma, g, s\}$, which define our best-fit model. If $g = -1$, it implies a zero covariance at large scales. We fit two models: a full model with all parameters free, and a reduced model with $g = -1$, and the rest of the parameters free. The priors for the parameters are specified in Table~\ref{tab:prior}.

For the full model, we choose the error function as our fiducial functional form. The estimated parameters for both the full and reduced models are presented in Table~\ref{tab:cov_best_fit_param_full_model} and Table~\ref{tab:error_func_reduced}, respectively. Appendix \ref{sec:cov_robustness} provides robustness testing to determine the best-fit model for our covariance. In Table~\ref{tab:cov_best_fit_param_full_model}, we compare the model parameters, $\chi^2$ p-value, and the difference in Deviance Information Criterion (DIC) with our fiducial model using the candidate functions listed in Table~\ref{tab:cov_functional_form}. The error function generally outperforms other models when all parameters $\boldsymbol{\theta} \in \{\tau, \gamma, g, s \}$ are allowed to vary. Table~\ref{tab:error_func_reduced} shows that the DIC of the reduced error function ($g = -1$) marginally outperforms the full error function in most cases, along with the posterior constraints of parameters shown in Figure~\ref{fig:cov_cornerplot}.

\begin{table}
\centering
\ra{1}
    \begin{tabular}{lll} \toprule
    \multirow{2}{*}{Parameters} & \multicolumn{2}{c}{Priors} \\
    \cmidrule{2-3}
    & Full  & Reduced \\ \midrule
    $\tau$ &             Uniform (0, 10) & Uniform (0, 10) \\
    $\gamma$ &           Uniform (-5, 5) & Uniform (-5, 5) \\
    $g$ &                Uniform (-2, 1) &  Fixed at $g=-1$ \\
    $10^{-12}\times s$ &  Log-uniform (0.01, 10) & Log-uniform (0.01, 10)  \\
    \bottomrule
    \end{tabular}
\caption{Priors for the model. We introduce two sets of priors. In the "full" models, the parameters are given physical (i.e., $\tau>0$, $s>0$) but non-informative uniform or log-uniform priors. In the ``reduced'' case, assuming that Cov$(\Delta\Sigma,\ln N_{\rm gal} \mid M, z)=0$ at large scales, we restrict $g = -1$ while assigning the same set of priors to all other parameters.}
\label{tab:prior}
\end{table}

\begin{table}
    \centering
    \begin{tabular}{|c|c|} \toprule
         \textbf{Error function (fiducial)} &  $s\Big( {\rm erf}$$(\frac{\sqrt{\pi}}{2} \tilde{x}) + g\Big)$ \\ \hline
         Logistics function & $s\Big(\frac{2}{1+e^{\tilde{x}}}-1  + g\Big)$ \\  \hline
         Inverse tangent & $s\Big(\frac{2}{\pi}\arctan\big(\frac{\pi}{2}\tilde{x}\big) + g \Big)$ \\ \hline
         Algebraic second order & $s(\tilde{x}/(1 + \tilde{x}^2)^{1/2} + g)$ \\ 
         \bottomrule
    \end{tabular}
    \caption{Functional forms to model Cov($\Delta\Sigma$, $\ln N_{\rm gal}$). The radius in log-space $x$ is transformed to $\tilde{x} \equiv (x-\gamma)/\tau$ by a horizontal offset $\gamma$ and a characteristic scale $\tau$. The functions $f(\tilde{x})$ are normalized so that $f(\tilde{x})$ asymptotically goes to 1 at $+\infty$, -1 at $-\infty$, $f(0)=0$ and $f'(0) = 1$. Finally, we wrap $f(\tilde{x})$ by the function $p(f(\tilde{x})) \equiv s(f(\tilde{x} +g)$ to include a vertical offset $g$ and amplitude parameter $s$. Together $\mathbf{\theta} \in \{\tau, \gamma, g, s \}$ form the set of model parameters that allow us to make apple-to-apple comparisons between models. }
    \label{tab:cov_functional_form}
\end{table}

\section{Particle Resolution and its impact on measurement errors} 
\label{sec:ptcl_resolution}
Using the 300 Cori Haswell node hours allocated by the NERSC to this project, we measured $\Delta\Sigma$ for $\sim 5000$ clusters using dark matter particles downsampled by a factor of 10, in 20 log-spaced radial bins, at a projection depth of $200~h^{-1}\rm Mpc$. 

At a downsampling rate of 10, our effective dark matter particle resolution is $M_{\rm p} \approx 1.51 \times 10^{10} h^{-1}M_\odot$. The error in $\Delta\Sigma$ comes from three sources: (i) cosmic variance, (ii) Poisson noise, and (iii) the intrinsic diversity of halos accounted for by secondary halo properties. In \S\ref{subsec: DS_Xparam}, we presented the contribution to $\Delta\Sigma$ scatter from secondary halo properties. Here, we compare the Poisson noise to the cosmic variance floor. 

The cosmic variance introduces fluctuations in a 2D surface density fluctuation, given by 
\beqa
\delta\Delta\Sigma(R) = D_p \rho_m \sigma(R),
\label{eqn:cosmic_2d_density_fluctuation}
\eeqa
where $D_{p} = 200 h^{-1}\rm Mpc$ is the projection depth, $\rho_m(z)$ is the mean density of the universe at that redshift, and $\sigma(R)$ is the root mean squared matter density fluctuation, given by 
\beqa
\sigma^2(R) = \int\Delta^2(k)\Big(\frac{3j_1(kR)}{kR}\Big)^2 d\ln{k},
\label{eqn:cosmic_variance}
\eeqa
which is smoothed over an area of $A=4\pi R^2$, $\Delta^2(k)$ is the matter power spectrum for a wavenumber $k$, and $j_1$ is the Bessel function of the first order. 

\begin{figure}
    \centering
    \includegraphics[width=\linewidth]{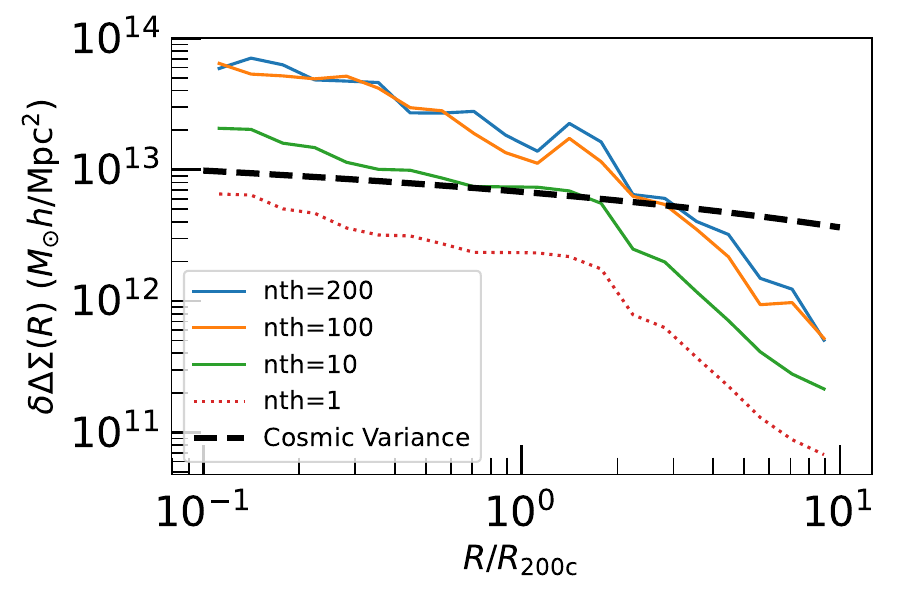}
    \caption{The standard error of $\Delta\Sigma$ measurements tested on a benchmark bin of $M_{\rm 200c} \in [1\times10^{14}, 2\times10^{14}) M_{\odot}h^{-1}$ at z=0. The standard error is estimated using the bootstrap method for the $N=500$ clusters with dark matter particles downsampled by a factor of 200, 100, and 10 (solid lines). At our current resolution (nth=10, solid green line), the standard error is just above the cosmic variance at small scales and drops below the cosmic variance at large scales. The solid black line is density fluctuation estimated from the cosmic variance floor, as described in Equations~\eqref{eqn:cosmic_variance} and \eqref{eqn:cosmic_2d_density_fluctuation}. In the ideal case that Poisson noise accounts for all the standard error, fully sampling all particles ($n{\rm th}=1$, red dotted line) will reduce the standard error by a factor of $\sqrt{10}$, rendering it just below the cosmic variance floor at small scales. In the realistic case that the standard error for $\Delta\Sigma$ contributes from both Poisson noise and the intrinsic diversity of halo profiles, the fully sampled standard error should be on par with the cosmic variance at small scales.}
    \label{fig:ptcl_resolution}
\end{figure}

Figure~\ref{fig:ptcl_resolution} shows the standard error of $\Delta\Sigma$ at a benchmark bin $M_{\rm 200c} \in [1\times10^{14}, 2\times10^{14}) M_{\odot}h^{-1}$ at z=0.00. At particle downsampling factors of 200, 100, and 10, the reduction in error is consistent with the Poisson term of $\sqrt{N}$, indicating that at these sampling rates, Poisson noise dominates. At our current downsampling rate of nth=10, the standard error is just above the cosmic variance floor at small scales and drops below the cosmic variance floor at large scales. In the ideal case that Poisson noise accounts for all the standard error, fully sampling all particles (nth=1, red dotted line) will reduce the standard error by a factor of $\sqrt{10}$, rendering it just below the cosmic variance floor at small scales. In the realistic case that the standard error for $\Delta\Sigma$ contributes from both Poisson noise and the intrinsic diversity of halo profiles, the fully sampled standard error should be on par with the cosmic variance at small scales. A future study with fully sampled particles should yield greater statistical constraints. 

\section{Derivation of Second Order Expansion Around the HMF}
\label{sec:derive_2nd_order}
Following the formalism from \citet{Evrard:2014} we derive Equation \ref{eq:2nd_order_correction_s_b}. The mean observable-mass scaling relation is given by the expression
\begin{equation}
    \langle s_i \mid \ln{M_0} \rangle = \alpha_i \ln{M_0} + \pi_i, 
    \label{eqn:mean_observable_mass_generic}
\end{equation}
for $s_i \in \{\DSig, \ln\Ngal\}$ and pivot mass $M_0$. We now denote the deviation from the mean relation as $\delta_i$, which from rearranging the terms in Equation \eqref{eqn:mean_observable_mass_generic} is given by $\delta_i = \frac{(s_i-\pi_i)}{\alpha_i} - \ln{M_0}$. From \citet{Evrard:2014} the expression for the observable scaling relation for generic observables $\{a,b\}$ that follow a log-linear scaling relation as in Equation \eqref{eqn:mean_observable_mass_generic} is given by 
\begin{equation}
    \langle \delta_b \mid s_a \rangle = x_a\big[ \langle \ln{M} \mid s_a \rangle + (\gamma_1 + \gamma_2 \delta_a) r_{ab}\sigma_{\ln{M}|a,1}\sigma_{\ln{M}|b,1} \big],
    \label{eqn:2nd_order_scaling_generic}
\end{equation}
where $\gamma_1$ and $\gamma_2$ are the first and second order coefficients of the Taylor expansion of the HMF around the pivot mass $M_0$ and $x_{a} = (1 + \gamma_2\sigma^2_{\ln{M}|a,1})^{-1}$ the curvature term. The subscript 1 denotes the scatter for the HMF expanded to first order. 

We now convert the left hand side of Equation \eqref{eqn:2nd_order_scaling_generic} from the deviation from the mean scaling relation, $\delta_b$, to the observable $s_b$ to arrive at the expression
\begin{align}
\langle s_b \mid s_a \rangle =& \big[\alpha_b x_a \langle\ln{M} \mid s_a \rangle + \ln{M_0} - \pi_b \big] + \nonumber  \\ 
&\big[\alpha_b x_a (\gamma_1 + \gamma_2\delta_a)r_{ab} \big]\sigma_{\lnM|a,1}\sigma_{\lnM|b,1} \nonumber \\ 
=&
\langle b \mid a, z \rangle_{\rm fid} + \nonumber \\ 
&{\rm Cov} (a, b) \times 
\Big[\frac{x_a}{\alpha^2_{a}} (\alpha_{a} \gamma_1 + \gamma_2(s_a - \pi_{a}))  \Big], 
\end{align}
where we made use of the fact $\text{Cov}(a,b|M,z) = r_{ab}\sigma_{a|M}\sigma_{b|M}$ and that the mass scatter conditioned on the observable to first order is related to the observable scatter by $\sigma_{\ln{M}|a,1} = \sigma_{a|M}/\alpha_{a}$, as shown in Equation 4 in \citet{Evrard:2014} for the multivariate case. Substituting $\ln\Ngal$ for $a$ and $\DSig$ for $b$ yields the expression for Equation \eqref{eq:2nd_order_correction_s_b}.

\section{Robustness testing of Covariance Modeling}
\label{sec:cov_robustness}
The shape posterior is sampled by a Monto Carlo Markov Chain (MCMC) using the emcee package \citep{Foreman_Mackey_2013}. We test for convergence by ensuring that the number of steps exceeds $100t_{\rm auto}$ for all parameters, where $t_{\rm auto}$ is the integrated autocorrelation time as defined by \citet{Goodman_Weare_2010} and by ensuring that the convergence diagnostic denoted with $R$ \citep{Gelman_Rubin_92} across all walkers satisfy $R < 1.05$. 

The posterior distribution according to Bayes theorem is given as:
\begin{align}
    p(\boldsymbol{\theta}|\{y_{i}\}) &\propto p({\{y_{i}\}|\boldsymbol{\theta}})p({\boldsymbol{\theta}}) \\
    &= \prod_{i} p({y_{i}|\boldsymbol{\theta}})p({\boldsymbol{\theta}}),
\end{align}
where the second line assumes independent and identical distribution ($\mathrm{i.i.d}$) for the data vectors. We set uniform priors $p({\boldsymbol{\theta}})$ shown in Table ~\ref{tab:prior} with signs and ranges motivated by the shape of the covariance (i.e., a negative $\gamma$ and positive $\tau$ to offset $f(\tilde{x})$ to the left and a positive $s$ and negative $g$ shifts the fitted curve downwards). 

We measure the goodness of fit using the left-tail $p$-value for the $\chi^2$ with $N_{\rm data} - N_{\rm dim} = 20 - 4 = 16$ degrees of freedom. We compare between models by the Deviance Information Criterion defined as
\begin{equation}
    \mathrm{DIC} = 2\overline{ D(\boldsymbol{\theta})} - D(\boldsymbol{\overline{\theta}}),
\end{equation}
where $\overline{\theta}$ is the best-fit parameters, and $D(\boldsymbol{\theta})$ is defined as
\begin{equation}
    D(\boldsymbol{\theta}) \equiv -2\log{(P({\{x_{i}\}|\boldsymbol{\theta}}))}.
\end{equation}
The performance between different functional forms (Table \ref{tab:cov_functional_form}) is reported in Table~\ref{tab:prior}.

The summary statistics for the posterior distribution of the covariance models are listed in Table~\ref{tab:cov_best_fit_param_full_model} as plotted against measurements in Figure~\ref{fig:cov_function_compare}. Among the functions, the error function has either a better or comparable fit to all other functions in all other bins, as indicated by their DIC parameters. In two bins $M_{\rm 200c} \in [5\times10^{14},1\times10^{15})$ at $z=0.49$ and $M_{\rm 200c} \in [2\times10^{14},5\times10^{15})$ at $z=1.03$, the amplitude of the covariance is too small relative to their errors for shape parameters to be well-constrained. The right tail $p$-value for $\chi^2$ is $p > 0.05$ for all but one bin. For this reason, we take the full error function as the nominal functional form. 

\begin{figure}
    \centering
    \includegraphics[width=\linewidth]{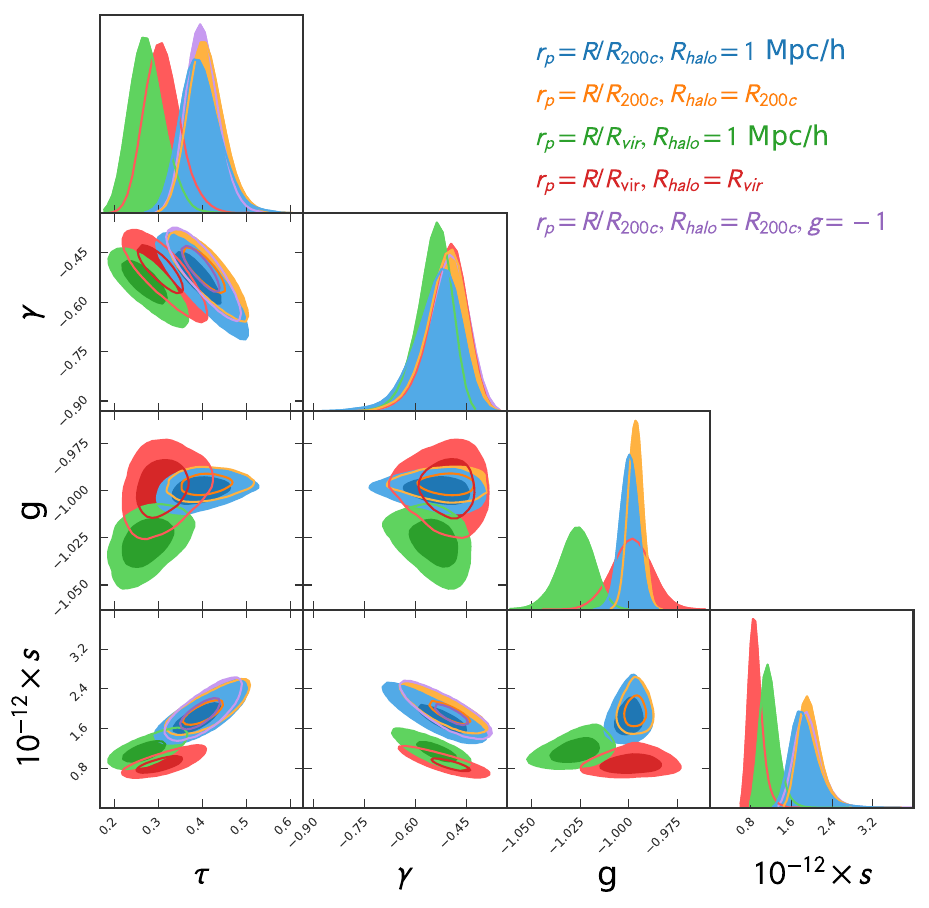}
    \caption{Posterior distribution of shape parameters in a benchmark bin of $M_{\rm 200c}/M_{\rm vir} \in [2\times10^{14},5\times10^{14})$ at $z=0.00$ under different binning schemes $r_p$ and $N_{\rm gal}$ models with different halo boundaries $R_{\rm halo}$. The marginalized parameter constraints for the full model closely overlap one another, and using the reduced model with $g=-1$ marginally improves the posterior constraints. The plot was generated using pygtc \citep{pygtc}.}
    \label{fig:cov_cornerplot}
\end{figure}

For $R \geq R_{\rm vir}$ or $R \geq R_{\rm 200c}$, we find the covariance to be null at $p$-values $> 0.01$. A zero covariance at large scales implies $g=-1$ which coincides with the reduced model. We compare the results of the error function of the reduced model to the full model and find their performance varies from bin to bin as indicated by the DIC (Table ~\ref{tab:error_func_reduced}). The posteriors of the reduced model provide marginally tighter constraints than the full model (Figure~\ref{fig:cov_cornerplot}). 

\begin{table*}
\centering
\ra{1.3}
    \begin{tabular}{lllllllll} \toprule
    \multirow{2}{*}{Mass \& redshift} & \multicolumn{8}{c}{Error function (full): $s\Big( {\rm erf}(\frac{\sqrt{\pi}}{2} \tilde{x}) + g\Big)$} \\
    \cmidrule{2-9}
    & $\tau$ & $\gamma$  & $g$ & $10^{12}\times s$ & $\Delta \rm DIC_{\rm log}$ & $\Delta \rm DIC_{\rm alg}$ & $\Delta \rm DIC_{\rm arctan}$ & $p$-value  \\ \midrule
    $[5\times10^{13},1 \times 10^{14})$, $z=0.00$ & $0.47^{+0.06}_{-0.05}$ & $-0.66^{+0.08}_{-0.11}$ & $-0.99^{+0.005}_{+0.004}$ & $2.88^{+0.71}_{-0.47}$ & 4.8 & 17.8 & 31.5 & 0.27 \\
    $[1\times10^{14},2\times10^{14})$, $z=0.00$ & $0.57^{+0.11}_{-0.08}$ & $-0.76^{+0.14}_{-0.22}$ & $-1.008^{+0.0005}_{-0.0007}$ & $2.73^{+0.13}_{-0.65}$ & -2.9 & -1.7 & 2.8 & 0.88 \\
    $[2\times10^{14},5\times10^{14})$, $z=0.00$ & $0.41^{+0.04}_{-0.04}$ & $-0.51^{+0.05}_{-0.06}$ & $-0.997^{+0.004}_{-0.0004}$ & $1.93^{+0.24}_{-0.18}$ & 4.4 & 17.4 & 35.0 & 0.79 \\
    $[5\times10^{14},1\times10^{15})$, $z=0.00$ & $0.27^{+0.03}_{-0.03}$ & $-0.40^{+0.04}_{-0.05}$ & $-1.006^{+0.005}_{-0.0005}$ & $1.03^{+0.13}_{-0.11}$ & -0.27 & 14.3 & 29.1 & 0.0009 \\
    \hline
    $[5\times10^{13},1\times10^{14})$, $z=0.49$ & $0.28^{+0.6}_{-0.05}$ & $-0.36^{+0.05}_{-0.07}$ & $-0.990^{+0.013}_{-0.013}$ & $0.75^{+0.11}_{-0.09}$ & 2.7 & 9.0 & 15.1 & 0.72 \\
    $[1\times10^{14},2\times10^{14})$, $z=0.49$ &  $0.35^{+0.07}_{-0.06}$ & $-0.43^{+0.07}_{-0.09}$ & $-1.022^{+0.016}_{-0.017}$ & $0.81^{+0.15}_{-0.10}$ & 1.7 & 5.4 & 9.3 & 0.97 \\
    $[2\times10^{14},5\times10^{14})$, $z=0.49$ & $0.24^{+0.08}_{-0.07}$ & $-0.49^{+0.07}_{-0.10}$ & $-0.995^{+0.012}_{-0.011}$ & $0.51^{+0.14}_{-0.01}$ & 0.6 & 2.4 & 4.0 & 0.63 \\
    $[5\times10^{14},1\times10^{15})$, $z=0.49$ & \multicolumn{7}{c}{NA} \\
    \hline
    $[5\times10^{13},1\times10^{14})$, $z=1.03$ & $0.17^{+0.05}_{-0.04}$ & $-0.31^{+0.05}_{-0.05}$ & $-1.019^{+0.020}_{-0.022}$ & $0.35^{+0.06}_{-0.05}$ & 4.3 & 6.5 & 8.3 & 0.05 \\
    $[1\times10^{14},2\times10^{14})$, $z=1.03$ & $0.21^{+0.06}_{-0.05}$ & $-0.42^{+0.05}_{-0.06}$ & $-1.024^{+0.013}_{-0.0014}$ & $0.51^{+0.09}_{-0.07}$ & -0.3 & 3.3 & 6.1 & 0.01 \\
    $[2\times10^{14},5\times10^{14})$, $z=1.03$ & \multicolumn{7}{c}{NA} \\
    
    \bottomrule
    \end{tabular}
\caption{Summary statistics for Cov($\Delta\Sigma$, $N_{\rm gal} \mid M, z$) binned by $R_{\rm 200c}$ and $M_{200c}$, with $N_{\rm gal}$ defined inside the halo $R_{\rm 200c}$. Columns 2-5 are the best-fit parameters for the nominal error function and their $1\sigma$ ranges. Columns 6-8 are the difference between the DIC of the logistics, algebraic, and inverse tangent models with the nominal error function, respectively. Column 9 is the right-tail p-value as measured by the $\chi^2$ statistic with 20-4 = 16 degrees of freedom. Across all bins with applicable posterior constraints, the error function out-performs or is comparable to alternative functional forms as indicated by the difference in DIC, and has $p \geq 0.01$ in all but one bin. In two bins $M_{\rm 200c} \in [5\times10^{14},1\times10^{15})$ at $z=0.49$ and $M_{\rm 200c} \in [2\times10^{14},5\times10^{15})$ at $z=1.03$ the size of the covariance is too small relative to the size of their errors for shape parameters to be constrained. The covariance in these two bins is consistent with null at $p =0.01$ and $p=0.05$ levels, respectively.
}
\label{tab:cov_best_fit_param_full_model}
\end{table*}

\begin{table*}
\centering
\ra{1.3}
    \begin{tabular}{cccccc} \toprule
    \multirow{2}{*}{Mass \& redshift} & \multicolumn{5}{c}{Error function (reduced): $s\Big( {\rm erf}(\frac{\sqrt{\pi}}{2} \tilde{x}) -1 \Big)$} \\
    \cmidrule{2-6}
    & $\tau$ & $\gamma$  & $10^{12}\times s$ & $\Delta \rm DIC_{\rm erf-full}$  & $p$-value  \\ \midrule
    $[5\times10^{13},1 \times 10^{14})$, $z=0.00$ & $0.44^{+0.04}_{-0.04}$ & $-0.63^{+0.07}_{-0.09}$  & $2.66^{+0.56}_{-0.39}$ & -5.2 & 0.06 \\
    $[1\times10^{14},2\times10^{14})$, $z=0.00$ & $0.69^{+0.11}_{-0.09}$ & $-0.98^{+0.20}_{-0.27}$  & $4.10^{2.31}_{1.19}$ & 4.5 & 0.96 \\
    $[2\times10^{14},5\times10^{14})$, $z=0.00$ & $0.40^{+0.04}_{-0.03}$ & $-0.50^{+0.05}_{-0.06}$  & $1.90^{+0.26}_{-0.20}$ & 0.2 & 0.78 \\
    $[5\times10^{14},1\times10^{15})$, $z=0.00$ & $0.29^{+0.04}_{-0.04}$ & $-0.40^{+0.04}_{-0.04}$  & $1.12^{+0.13}_{-0.11}$ & 6.1 & 0.005 \\
    \hline
    $[5\times10^{13},1\times10^{14})$, $z=0.49$ & $0.25^{+0.04}_{-0.03}$ & $-0.35^{+0.04}_{-0.04}$  & $0.79^{+0.08}_{-0.07}$ & 6.2 & 0.57 \\
    $[1\times10^{14},2\times10^{14})$, $z=0.49$ & $0.38^{+0.14}_{-0.09}$ & $-0.48^{+0.12}_{-0.24}$  & $0.85^{+0.46}_{-0.19}$ & -6.9 & 0.81 \\
    $[2\times10^{14},5\times10^{14})$, $z=0.49$ & $0.25^{+0.21}_{-0.12}$ & $-0.55^{+0.14}_{-0.42}$  & $0.40^{+0.69}_{-0.13}$ & NA & 0.48 \\
    $[5\times10^{14},1\times10^{15})$, $z=0.49$ & $0.61^{+0.27}_{-0.24}$ & $-0.90^{+0.49}_{-0.76}$  & $1.01^{+3.05}_{-0.60}$ & -5.7 & 0.99 \\
    \hline
    $[5\times10^{13},1\times10^{14})$, $z=1.03$ & $0.20^{+0.06}_{-0.05}$ & $-0.35^{+0.06}_{-0.08}$  & $0.38^{+0.07}_{-0.06}$ & -0.75 & 0.04 \\
    $[1\times10^{14},2\times10^{14})$, $z=1.03$ & $0.22^{+0.06}_{-0.05}$ & $-0.42^{+0.06}_{-0.8}$  & $0.54^{+0.11}_{-0.09}$ & 3.9 & 0.3 \\
    $[2\times10^{14},5\times10^{14})$, $z=1.03$ & $5.97^{+2.76}_{-3.01}$ & $-0.55^{+3.30}_{-2.99}$  & $0.02^{+0.02}_{-0.008}$ & NA & 0.15 \\
    
    \bottomrule
    \end{tabular}
\caption{Summary statistics for Cov($\Delta\Sigma, N_{\rm gal} \mid M, z$) binned by $R_{\rm 200c}$ and $M_{200c}$ with $N_{\rm gal}$ defined inside the halo $R_{\rm 200c}$ for the reduced error function model. Compared with the full error function model, the performance of the reduced model varies from bin to bin -- using $\Delta{\rm DIC} > 3$ as a statistically significant result, it outperforms the full model in 4/9 overlapping bins, under-performs in 3/9 bins, and is comparable in 2 bins. The reduced model is able to yield convergent chains for $M_{\rm 200c} \in [5\times10^{14},1\times10^{15})$ at $z=0.49$ and $M_{\rm 200c} \in [2\times10^{14},5\times10^{15})$ at $z=1.03$ but with poor constraints on the parameters. }
\label{tab:error_func_reduced}
\end{table*}


\section{Modeling Secondary Properties}
\label{sec:multilinear_modeling}

We describe the linear regression model for richness used in \S\ref{subsec:richness_Xparam}. The same methodology is applied to $\Delta\Sigma$ in \S\ref{subsec: DS_Xparam}. 

To model the expected natural logarithm of galaxy count ($\ln\Ngal$), we decompose it linearly using secondary halo parameters listed in Table~\ref{tab:notation_Xparam}, as shown in Equation~\eqref{eqn:model_richness_Xparam}. We employ the least squares method for linear regression and examine parameter redundancies. For over half of the bins, the parameters $\Gamma_{\rm inst}$, $\Gamma_{\rm 100Myr}$, $\Gamma_{\rm 2dyn}$, and $\Gamma_{\rm peak}$ exhibit collinearity, with Variance Inflation Factors (VIF) exceeding 5. This outcome is expected, as these quantities represent the same physical quantities smoothed over different time scales. As for the reduced set of non-collinear parameters, their correlation coefficient are quantified in \citet{Shin_2023} using the Erebos simulation suite. To determine which parameters to retain, we utilize the partial F-statistic.

Table~\ref{tab:richness_multilinear_coeff} demonstrates the diminishing explanatory power of $\Pi$ on the richness as seen by the diminishing $R^2$ and $F_{\rm partial}$. We consider the partial F-statistic serves as a heuristic measure for the explanatory power of a variable, defined as:
\beqa
F_{\rm partial} = \frac{({\rm RSSE_{\rm reduced} - RSSE_{\rm full}})/p}{{\rm RSSE_{\rm full}}/(n-k)},
\label{eqn:F_partial}
\eeqa
where RSSE is the residual sum of squared errors for the reduced model after removing the parameter in question and the full model containing $\Pi \subset ~\{a_{1/2}, c_{\rm vir}, T/U, \Gamma_{\rm 2dyn}, X_{\rm off}\}$, $p$ is the number of parameters removed from the full model which in our case is by construction set to $p = 1$, $n$ the number of data points, and $k$ is the number of parameters in the full backward model. This statistic can be shown to be proportional to the contribution to the total $R^2$ \textit{uniquely} explained by this parameter alone.  

A partial F-statistic test reveals that $\Gamma_{\rm 2dyn}$ exhibits the highest partial F-statistic of all accretion rate parameters. Therefore, we retain this parameter in the reduced dimensional linear regression. The final model includes the following parameters $\Pi \subset \{a_{1/2}, c_{\rm vir}, T/U, \Gamma_{\rm 2dyn}, X_{\rm off}\}$. To ensure the robustness of the linear model across all bins, we perform the following tests:
\begin{itemize}
    \item Variance inflation factor (VIF) test with a cutoff of 5 to detect multicollinearity.
    \item Global F-statistic for the entire model with a significance level of $0.05$ to examine the correlation between the dependent variable and all parameters.
    \item Partial F-statistic for the entire parameter set to compare the relative importance of each parameter. The Partial F-statistic measures the additional contribution of each parameter to the multi-linear fit by estimating its corresponding $R^2$ value. 
    \item T-statistic for each parameter to verify that the coefficients significantly deviate from zero at a significance level of $0.05$.
    \item Breusch-Pagan Lagrange Multiplier test \citep{Breusch_Pagan_1979} at a significance level of $0.05$ to assess heteroscedasticity.
    \item Shapiro-Wilk test \citep{Shapiro_Wilk_1965} at a significance level of $0.05$ to evaluate the Gaussianity of residuals.
\end{itemize}
Across all bins, the reduced model successfully satisfies the first four tests. However, some bins fail the Shapiro-Wilk test due to a negative skew and positive kurtosis. Nonetheless, a visual examination of Q-Q plots indicates that the residuals predominantly follow a Gaussian distribution, except for deviations at the tail ends. Q-Q plots, quantile-quantile plots, are visualization tools used to compare the quantiles of a dataset to the quantiles of a theoretical distribution, typically a normal distribution. They provide a visual assessment of how well the data aligns with the assumed distribution. 

\begin{table*}
\centering
\ra{1.3}
    \begin{tabular}{lcccccccccccc} \toprule
    \multirow{2}{*}{Redshift} & \multirow{2}{*}{$R^2$} &  \multirow{2}{*}{Const} & \multicolumn{2}{c}{$c_{\rm vir}$} & \multicolumn{2}{c}{T/|U|}  & \multicolumn{2}{c}{$a_{1/2}$} & \multicolumn{2}{c}{$\Gamma^*_{\rm 2dyn}$}  & \multicolumn{2}{c}{$X_{\rm off}$} \\
    \cmidrule{4-13}
    \& $M_{\rm 200c}$ ($M_{\odot}h^{-1}$) &  & & $\beta_{\Ngal}$ & $F$ & $\beta_{\Ngal,}$ & $F$ & $\beta_{\Ngal}$ & $F$ &
    $10^6\times \beta_{\Ngal}$ & $F$ & $10^3\times\beta_{\Ngal}$ & $F$ 

    \\ \midrule
    $z=0.00$  &&&&&&&&&&&& \\
    \multirow[t]{2}{*}{$[5\times10^{13}, 1\times10^{14})$} & 0.45 & -0.69 & -0.035(0.004) & 171 & 1.448(0.135) & 185 & 0.103(0.084) & 166 & 2.743(1.68) & 134 & -1.62(0.23) & 69  \\
    
    \multirow[t]{2}{*}{$[1\times10^{14}, 2\times10^{14})$} & 0.48 & -0.46 & -0.033(0.003) & 217 & 1.086(0.120) & 179 & -0.198(0.071) & 170 & 2.668(0.628) & 206 & -0.76(0.15) & 86  \\
        
    \multirow[t]{2}{*}{$[2\times10^{14}, 5\times10^{14})$} & 0.49 & -0.41 & -0.029(0.003) & 196 & -0.794(0.086) & 145 & 0.292(0.007) & 213 & -0.722(0.223) & 167 & 0.57(0.11) & 67  \\
        
    \multirow[t]{2}{*}{$[5\times10^{14}, 1\times10^{15})$} & 0.46 & -0.32 & -0.019(0.002) & 119 & 1.503(0.068) & 111 & 0.148(0.054) & 201 & 0.483(0.096) & 226 & -0.25(0.06) & 51  \\
    \hline
    
    $z=0.49$  &&&&&&&&&&&& \\
    \multirow[t]{2}{*}{$[5\times10^{13}, 1\times10^{14})$} &0.25& -0.71 & 0.006(0.005) & 24 & 0.325(0.125) & 69 & 0.188(0.101) & 142 & 3.90(1.64) & 102 & -0.3(0.19) & 35  \\
    
    \multirow[t]{2}{*}{$[1\times10^{14}, 2\times10^{14})$} & 0.21 & -0.45 & -0.450(0.082) & 27 & 0.314(0.135) & 46 & -0.353(0.158) & 107 & 1.80(0.71) & 89 & -0.37(0.13) & 14  \\
       
    \multirow[t]{2}{*}{$[2\times10^{14}, 5\times10^{14})$} & 0.13 & -0.56 & -0.003(0.004) & 0 & 0.323(0.110) & 26 & 0.670(0.155) & 51 & 0.388(0.108) & 26 & -0.27(0.11) & 2  \\
       
    \multirow[t]{2}{*}{$[5\times10^{14}, 1\times10^{15})$} & 0.13 & -0.32 & -0.010(.008) & 0 & 0.007(0.286) & 0 & 0.697(.309) & 5 & -0.538(0.331) & 3 & -0.41(.14) & 3  \\
    \hline
    
    $z=1.03$  &&&&&&&&&&&& \\
    \multirow[t]{2}{*}{$[5\times10^{13}, 1\times10^{14})$} & 0.27 & -1.12 & 0.015(0.005) & 6 & 0.471(0.136) & 21 & 0.569(0.269)& 116 & 2.7(1.2) & 63 & -0.81(0.18) & 0  \\
    
    \multirow[t]{2}{*}{$[1\times10^{14}, 2\times10^{14})$} & 0.22 & -0.84 & 0.016(0.004) & 10 & 0.374(0.135) & 36 & -0.233(0.234) & 83 & 1.076(0.478) & 36 & -0.35(0.12) & 2  \\
        
    \multirow[t]{2}{*}{$[2\times10^{14}, 5\times10^{14})$} & 0.20 & -0.6 & 0.0015(0.005) & 10 & 0.284(0.185) & 1 &  -0.157(0.266) & 15 & 0.160(0.213) & 6 & -0.55(0.11)
    & 13  \\
        
    \bottomrule
    \end{tabular}
\caption{Best-fit parameters, global $R^2$, and explanatory power indicators for log-richness modeled in Equation~\eqref{eqn:model_richness_Xparam}. Values in parentheses represent $1\sigma$ confidence intervals to the partial slopes $\beta$. The partial $F$-statistic, defined in Equation~\eqref{eqn:F_partial}, is used to quantify the explanatory power of each variable. A higher partial F-statistic indicates a greater amount of predictive power uniquely attributed to that variable. Statistical significance is determined by an F-statistic of $F>10$.} 
\label{tab:richness_multilinear_coeff}
\end{table*}

\section{Derivation of $P(\Delta\Sigma \mid \Ngal, m, z)$} 
\label{sec:proof_DS_given_lambda}
We demonstrate that $P(\Delta\Sigma \mid \Ngal, M, z)$ can be modeled as a multi-linear relation of secondary halo parameters of mean and variance given by Equations \eqref{eq:DS_given_lambda_Xparam_mean} and \eqref{eq:DS_given_lambda_Xparam_var}.

From our assertion that $P(\Delta\Sigma, \vec{\Pi} \mid M, z)$ is a bivariate normal, the conditional probability $P(\Pi_i \mid \Delta\Sigma, M, z)$ in each radial bin can be expressed as a normal distribution with mean
\beq
\avg{\Pi_i|\Delta\Sigma,  M, z} = \avg{\Pi_i | M, z}+ \rho_{\Pi_i-\Delta\Sigma}\frac{\sigma_{\Pi_i}}{\sigma_{\Delta\Sigma}} (\Delta\Sigma - \avg{\Delta\Sigma | M, z}),
\label{eq:Pi_given_DS_mean}
\eeq
and variance
\beq
\sigma_{\Pi_i | \Delta\Sigma}^2 = \sigma^2_{\Pi_i}(1-\rho^2_{\Pi_i-\Delta\Sigma}). 
\label{eq:Pi_given_DS_var}
\eeq
Here, we omit the radial dependence $R/R_{\rm 200c}$ for all variables and the conditional dependence on $(M, z)$ in the subscripts for $\rho$ and $\sigma$ (i.e., $\sigma_{\Delta\Sigma}$ should be explicitly written as $\sigma_{\Delta\Sigma| M, z}(R)$). 

For an independent random variable $Z = X+Y$ with $X$ and $Y$ uncorrelated independent random variables with distributions of the form $\sim \mathcal{N}(u,\sigma^2)$ and $\sim \mathcal{N}(\nu,\tau^2)$, respectively, Z is another Gaussian of $\mathcal{N}(u+v, \sigma^2 + \tau^2)$. In the case that X and Y are correlated, we must introduce a cross term in the variance of probability distribution Z, namely $P(Z) \sim \mathcal{N}(u+v, \sigma^2 + 2\rho_{X-Y}\sigma\tau + \tau^2)$. 

In our specific case, we want to model the distribution $P(\sum_i \beta_{s,i}\Pi_i \mid \Delta\Sigma, M, z)$ from $P(\Pi_i \mid \Delta\Sigma, M, z)$ and the correlation here refers to the correction between secondary halo parameter $\rho_{\Pi_i, \Pi_j}$.\footnote{In \S\ref{subsec: DS_Xparam}, we show through the Variance Inflation test that set of reduced model parameters $\Pi \in \{a_{1/2}, c_{\rm vir}, T/U, \Gamma_{\rm 2dyn}^*, X_{\rm off} \}$ are not multi-collinear.} From the convolution theorem and the expressions for $P(\Pi_i \mid \Delta\Sigma, M, z)$ in Equations~\eqref{eq:Pi_given_DS_mean} and \eqref{eq:Pi_given_DS_var}, we obtain the expression for $P(\ln\Ngal \mid \Delta\Sigma, M, z)$ as a normal distribution with mean:
\begin{align}
\avg{\ln\Ngal | \Delta\Sigma, M, z } ~=&~ \avg{\ln N_{\rm gal,0}|\Delta\Sigma, M, z}+ \\ \nonumber 
& \sigma_{\Delta\Sigma}\Big(\sum_i{\frac{\beta_{\Ngal,i}}{\sigma_{\Pi_i}}\rho_{\Delta\Sigma-\Pi_i}\times}(\Pi_i - \avg{\Pi_i | M, z})\Big),
\end{align}
and variance
\begin{align}
\qquad \sigma^2_{\ln{\Ngal}|\Delta\Sigma} \quad =&~  \sigma^2_{N_{\rm gal,0}} +  \sum_i \beta_{\Ngal,i}^2 \sigma^2_{\Pi_i}(1-\rho^2_{\Delta\Sigma-\Pi_i}) + \\ \nonumber
& \sum^{j\neq i}_{i,j}\rho_{\Pi_i-\Pi_j}\sigma_{\Pi_i}\sigma_{\Pi_j}.
\end{align}
We now want to derive the scaling relations for $P(\Delta\Sigma \mid \Ngal, M, z)$. From the Bayes theorem,
\beqa
P(\Delta\Sigma \mid \Ngal, M, z) = P(\ln\Ngal \mid \Delta\Sigma, M, z) \frac{P{(\Delta\Sigma \mid M, z)}}{P(\ln\Ngal \mid M, z)},
\label{eq:P_DS_Bayes}
\eeqa
wherein we assume that $P(\ln\Ngal | M, z)$ and $P(\Delta\Sigma | M, z)$ can be modeled as normal distributions, then  $P(\Delta\Sigma \mid \Ngal, M, z)$ is another normal distribution with mean 
\begin{align}
\label{eq:DS_given_lambda_Xparam_mean_append}
\avg{\Delta\Sigma | \Ngal, M, z } ~=&~ \avg{\Delta\Sigma|N_{\rm gal,0}, M, z} \\ \nonumber 
+& C_1\sigma_{\Delta\Sigma}\Big(\sum_i \frac{\beta_{\Ngal,i} }{\sigma_{\Pi_i}}\rho_{\Delta\Sigma-\Pi_i} \times (\Pi_i - \avg{\Pi_i | M, z})\Big)
\end{align}
and variance
\begin{align}
\label{eq:DS_given_lambda_Xparam_var_append}
\qquad \sigma^2_{\Delta\Sigma|\ln{\Ngal}} \quad =&~  \sigma^2_{0} +  C_2\sum_i{\beta_{\Ngal,i}^2 \sigma^2_{\Pi_i}(1-\rho^2_{\Delta\Sigma-\Pi_i})}+ \\ \nonumber
& C_3\sum^{j\neq i}_{i,j}{\rho_{\Pi_i-\Pi_j}\sigma_{\Pi_i}\sigma_{\Pi_j}}.
\end{align}
The parameter $C_1$ for the mean relation can be explicitly derived
if we know the posterior distribution of  $P(\ln\Ngal | M, z)$ and $P(\Delta\Sigma | M, z)$ by the exercise of completing the squares inside the exponents, i.e. by matching the quadratic, linear and constant terms inside the exponents of normal distributions on the left and right hand sides of Equation \eqref{eq:P_DS_Bayes}.

The parameters $C_2$, $C_3$, and $\sigma_0$ can be explicitly derived first by using the variance of product law of correlated Gaussians, i.e.
Var($Z \equiv XY) = 1 + \rho^2$ after transforming $X$ and $Y$ into unit variance, zero mean Gaussians with correlation efficient $\rho$, and then by using the variance of quotient approximation
\beqa
\rm{Var(X/Y)} = \big(\frac{\mu_X}{\mu_Y}\big)^2\Big[ \frac{\sigma_R^2}{\mu_R^2} -2 \frac{\rm{Cov(X,Y)}}{\mu_X \mu_Y} + \frac{\sigma_X^2}{\mu_Y^2}\Big],
\eeqa
where $\mu_X$, $\sigma_X^2$ are the mean and variances of $P(\Delta\Sigma \mid M, z)$ and $\mu_Y$, $\sigma_Y^2$ the mean and variances of $P(\ln\Ngal \mid M, z)$ in our specific case. 

The exact values of $C1$, $C2$, $C3$, and $\sigma_0$ are not essential for this paper as we aim to derive a general expression for $P(\Delta\Sigma \mid \Ngal, M, z)$ as a function of secondary halo parameters.


\bsp	
\label{lastpage}
\end{document}